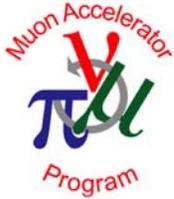

# U.S. Muon Accelerator Program

# Enabling Intensity and Energy Frontier Science with a Muon Accelerator Facility in the U.S.:
## *A White Paper Submitted to the 2013 U.S. Community Summer Study of the Division of Particles and Fields of the American Physical Society*

### *Contributed by the U.S. Muon Accelerator Program (MAP) and Associated Collaborators*

**Editors:**

**J-P. Delahaye (SLAC), C. Ankenbrandt (Muons Inc./FNAL), A. Bogacz (JLAB), S. Brice (FNAL), A. Bross (FNAL), D. Denisov (FNAL), E. Eichten (FNAL), P. Huber (VT), D.M. Kaplan (IIT), H. Kirk (BNL), R. Lipton (FNAL), D. Neuffer (FNAL), M.A. Palmer (FNAL), R. Palmer (BNL), R. Ryne (LBNL), P. Snopok (IIT/FNAL)**

*Abstract:*

A staged approach towards muon based facilities for Intensity and Energy Frontier science, building upon existing and proposed facilities at Fermilab, is presented. At each stage, a facility exploring new physics also provides an R&D platform to validate the technology needed for subsequent stages. The envisioned program begins with nuSTORM, a sensitive sterile neutrino search which also provides precision neutrino cross-section measurements while developing the technology of using and cooling muons. A staged Neutrino Factory based upon Project X, sending beams towards the Sanford Underground Research Facility (SURF), which will house the LBNE detector, could follow for detailed exploration of neutrino properties at the Intensity Frontier, while also establishing the technology of using intense bunched muon beams. The complex could then evolve towards Muon Colliders, starting at 126 GeV with measurements of the Higgs resonance to sub-MeV precision, and continuing to multi-TeV colliders for the exploration of physics beyond the Standard Model at the Energy Frontier. An Appendix addresses specific questions raised by the Lepton Colliders subgroup of the CSS2013 Frontier Capabilities Study Group.

# U.S. Muon Accelerator Program

## Table of Contents

# U.S. Muon Accelerator Program

## Executive Summary

Muon accelerators offer unique potential for the U.S. High Energy Physics community. In 2008, and subsequently in 2010, the U.S. Particle Physics Project Prioritization Panel (P5)[1,2] recommended that a world-class program of Intensity Frontier science be pursued at Fermilab as the Energy Frontier program based on the Tevatron reached its conclusion. Accordingly, Fermilab has embarked on the development of a next generation neutrino detector with LBNE and a next generation proton source with Project X. However, we must also consider what steps beyond those facilities would enable the continuation of a preeminent U.S. HEP research program. Building on the foundation of Project X, muon accelerators can provide that next step with a high intensity and precise source of neutrinos to support a world-leading research program in neutrino physics.

Moreover, the infrastructure developed to support such an Intensity Frontier research program can enable the return of the U.S. high energy physics program to the Energy Frontier: a subsequent stage of the facility could support one or more Muon Colliders, which could operate at center-of-mass energies from the Higgs resonance at 126 GeV up to the multi-TeV scale. Thus muon accelerators offer the unique potential, among the accelerator concepts being discussed for the Community Summer Study process, to provide world-leading experimental support spanning physics at both the Intensity and Energy Frontiers.

Before addressing the technical challenges of such facilities we summarize the cutting-edge physics they can do. For the proposed staging plan, baseline parameter specifications have been developed for a series of facilities, each capable of providing cutting-edge physics output, and at each of which the performance of systems required for the next stage can be reliably evaluated. The plan thus provides clear decision points before embarking upon each subsequent stage. The staging plan builds on the foundation of existing and proposed facilities, specifically:

- Project X at Fermilab as the megawatt-class proton driver for muon generation[3];
- Sanford Underground Research Facility (SURF), as developed for the LBNE detector. Neutrino Factory beams could initially be directed to an existing LBNE and ultimately to an upgraded detector that is optimized to take full advantage of those beams.

The performance characteristics of each stage provide unique physics reach:

- nuSTORM[4] (Neutrinos from STORed Muons): a short baseline Neutrino Factory (NF) enabling a definitive search for sterile neutrinos, as well as neutrino cross-section measurements that will ultimately be required for precision measurements at any long baseline experiment.
- NuMAX (Neutrinos from Muon Accelerators at Project X): an initial long baseline Neutrino Factory, optimized for a detector at SURF—a precise and well-characterized neutrino source that exceeds the capabilities of conventional superbeam technology.
- NuMAX+: a full intensity Neutrino Factory, upgraded from NuMAX, as the ultimate source to enable precision CP violation measurements in the neutrino sector.
- Higgs Factory: a collider whose baseline configurations are capable of providing between 3,500 and 13,500 Higgs events per year with exquisite energy resolution.
- Multi-TeV Collider: if warranted by LHC results, a multi-TeV Muon Collider (MC) likely offers the best performance and least cost for any lepton collider operating in the multi-TeV regime.

# U.S. Muon Accelerator Program

Nominal parameters for the three Neutrino Factories—the (short baseline) nuSTORM and two stages of (long baseline) NuMAX—are provided in Table 1. Collider parameters for a Higgs Factory as well as 1.5 and 3.0 TeV colliders are provided in Table 2. All of these machines would fit readily within the footprint of the Fermilab site. The ability to deploy these facilities in a staged fashion offers major benefits:

1. The strong synergies among the critical elements of the accelerator complex maximize the size of the experimental community that can be supported by the overall facility;
2. The staging plan reduces the investment required at each step to levels that will hopefully fit within the future budget profile of the U.S. high energy physics program.

**Table 1:** Muon Accelerator Program baseline Neutrino Factory parameters for nuSTORM and two NuMAX phases located on the Fermilab site and pointed towards a detector at SURF. For comparison, the parameters of the IDS-NF are also shown.

| System | Parameters | Unit | nuSTORM | NuMAX | NuMAX+ | IDS-NF |
|---|---|---|---|---|---|---|
| **Performance** | **Stored μ+ or μ-/year** | | $8\times10^{17}$ | $2\times10^{20}$ | $1.2\times10^{21}$ | $1\times10^{21}$ |
| | **$v_e$ or $v_\mu$ to detectors/yr** | | $3\times10^{17}$ | $8\times10^{19}$ | $5\times10^{20}$ | $5\times10^{20}$ |
| **Detector** | *Far Detector:* | Type | SuperBIND | MIND / Mag LAr | MIND / Mag LAr | MIND |
| | **Distance from Ring** | km | 1.9 | 1300 | 1300 | 2000 |
| | **Mass** | kT | 1.3 | 30 / 10 | 100 / 30 | 100 |
| | **Magnetic Field** | T | 2 | 0.5-2 | 0.5-2 | 1-2 |
| | *Near Detector:* | Type | SuperBIND | Suite | Suite | Suite |
| | **Distance from Ring** | m | 50 | 100 | 100 | 100 |
| | **Mass** | kT | 0.1 | 1 | 2.7 | 2.7 |
| | **Magnetic Field** | T | Yes | Yes | Yes | Yes |
| **Neutrino Ring** | **Ring Momentum ($P_\mu$)** | GeV/c | 3.8 | 5 | 5 | 10 |
| | **Circumference (C)** | m | 480 | 600 | 600 | 1190 |
| | **Straight section** | m | 185 | 235 | 235 | 470 |
| | **Arc Length** | m | 50 | 65 | 65 | 125 |
| **Acceleration** | **Initial Momentum** | GeV/c | - | 0.22 | 0.22 | 0.22 |
| | **Single-pass Linac** | GeV/pass | - | 0.95 | 0.95 | 0.56 |
| | | MHz | - | 325 | 325 | 201 |
| | **4.5-pass RLA**  RLA I | GeV/pass | - | 0.85 | 0.85 | 0.45 |
| | | MHz | - | 325 | 325 | 201 |
| | RLA II | GeV/pass | - | - | - | 1.6 |
| | | MHz | - | - | - | 201 |
| **Cooling** | | | No | No | 4D | 4D |
| **Proton Source** | **Proton Beam Power** | MW | 0.2 | 1 | 3 | 4 |
| | **Proton Beam Energy** | GeV | 120 | 3 | 3 | 10 |
| | **Protons/year** $1\times10^{21}$ | | 0.1 | 41 | 125 | 25 |
| | **Repetition Frequency** | Hz | 0.75 | 70 | 70 | 50 |

# U.S. Muon Accelerator Program

nuSTORM's capabilities could be deployed now. The NuMAX options and initial Higgs Factory could be based on the 3 GeV proton source of Project X Stage II operating with 1 MW and, eventually, 3 MW proton beams. This opens the possibility of launching the initial NuMAX, which requires no cooling of the muon beams, within the next decade. Similarly, the R&D required for a decision on a collider could be completed by the middle of the next decade. A Muon Collider in the multi-TeV range would offer exceptional performance due to the absence of synchrotron radiation effects, no beamstrahlung issues at the interaction point, and anticipated wall power requirements at the 200 MW scale, well below the widely accepted 300 MW maximum affordable power for a future HEP facility. This timeline, showing the targeted dates where critical decisions should be possible, is summarized in Figure 1.

**Table 2:** Muon Accelerator Program baseline Muon Collider parameters for both Higgs Factory and multi-TeV Energy Frontier colliders. An important feature of the staging plan is that collider activity could begin with Project X Stage II beam capabilities at Fermilab.

| Muon Collider Baseline Parameters | | | | | |
|---|---|---|---|---|---|
| | | **Higgs Factory** | | **Multi-TeV Baselines** | |
| *Parameter* | *Units* | Startup Operation | Production Operation | | |
| CoM Energy | TeV | 0.126 | 0.126 | 1.5 | 3.0 |
| Avg. Luminosity | $10^{34}$cm$^{-2}$s$^{-1}$ | 0.0017 | 0.008 | 1.25 | 4.4 |
| Beam Energy Spread | % | 0.003 | 0.004 | 0.1 | 0.1 |
| Higgs/$10^7$sec | | 3,500 | 13,500 | 37,500 | 200,000 |
| Circumference | km | 0.3 | 0.3 | 2.5 | 4.5 |
| No. of IPs | | 1 | 1 | 2 | 2 |
| Repetition Rate | Hz | 30 | 15 | 15 | 12 |
| $\beta^*$ | cm | 3.3 | 1.7 | 1 (0.5-2) | 0.5 (0.3-3) |
| No. muons/bunch | $10^{12}$ | 2 | 4 | 2 | 2 |
| No. bunches/beam | | 1 | 1 | 1 | 1 |
| Norm. Trans. Emittance, $\varepsilon_{TN}$ | π mm-rad | 0.4 | 0.2 | 0.025 | 0.025 |
| Norm. Long. Emittance, $\varepsilon_{LN}$ | π mm-rad | 1 | 1.5 | 70 | 70 |
| Bunch Length, $\sigma_s$ | cm | 5.6 | 6.3 | 1 | 0.5 |
| Beam Size @ IP | μm | 150 | 75 | 6 | 3 |
| Beam-beam Parameter / IP | | 0.005 | 0.02 | 0.09 | 0.09 |
| Proton Driver Power | MW | 4[#] | 4 | 4 | 4 |

[#] Could begin operation with Project X Stage 2 beam

# U.S. Muon Accelerator Program

The U.S. Muon Accelerator Program (MAP) has the task of assessing the feasibility of muon accelerators for Neutrino Factory and Muon Collider applications. Critical path R&D items important to the performance of one or more of these facilities include:

- Development of a high power target station which is ultimately capable of handling ≥4 MW of power. Liquid-metal jet technology has been shown to be capable of handling the necessary beam power[5]. While the complete engineering design of a multi-MW target station, including a high field capture solenoid (nominally 20 T hybrid normal and superconducting magnet with ~3 GJ stored energy) is challenging, target stations with similar specifications are required for other planned facilities (e.g., spallation sources), and our expectation is that the engineering challenges can be successfully addressed over the course of the next decade. In the meantime, a muon accelerator complex can begin producing world-class physics with the proton beam powers that will become available with Project X Stage II.

- Muon cooling is required in order to achieve the beam parameters for a high performance NF and for all MC designs under consideration. An ionization cooling channel requires the operation of RF cavities in tesla-scale magnetic fields. Promising recent results from the MuCool Test Area (MTA) at Fermilab point towards solutions to the breakdown problems of RF cavities operating in this environment[6]. These advances, along with technology concepts developed over the past decade, are expected to allow MAP to establish a baseline 6D cooling design on the 2-year timescale[7]. In addition, the Muon Ionization Cooling Experiment is expected to begin producing relevant results in the same time frame[8].

- High intensity, low energy beams (~200 MeV/c, optimal for muon ionization cooling) are susceptible to a range of potential collective effects. Evaluating the likely impact of these effects on the muon beams required for NF and MC applications, through simulation and experiment, is an important deliverable of the MAP feasibility assessment.

- For the MC, muon decays in the ring impact both the magnet and shielding design for the collider itself as well as backgrounds in the detector. Detector backgrounds have been shown to be manageable via pixelated detectors with good time resolution[9]. Thus, this issue appears to present no impediment to moving forward with full detector studies and machine–detector interface design efforts.

A thorough evaluation of these issues is crucial for an informed community decision on muon accelerator facilities. Furthermore, the proposed staging plan enables the performance, at each stage, of confirming R&D for the next stage in the plan, thus enabling a well-informed decision process moving forward.

To summarize, muon accelerators can enable a broad and world-leading high energy physics program which can be based on the infrastructure of the single remaining U.S high energy physics laboratory, Fermilab. While any decision to move forward with muon accelerator based technologies rests on the evolving physics requirements of the field, as well as the successful conclusion of the MAP feasibility assessment later this decade, the ability of muon accelerators to address crucial questions on both the Intensity and Energy Frontiers, as well as to provide a broad foundation for a vibrant U.S. HEP program, argues for a robust development program to continue. This will enable a set of informed decisions by the U.S. community starting near the end of this decade.

# U.S. Muon Accelerator Program

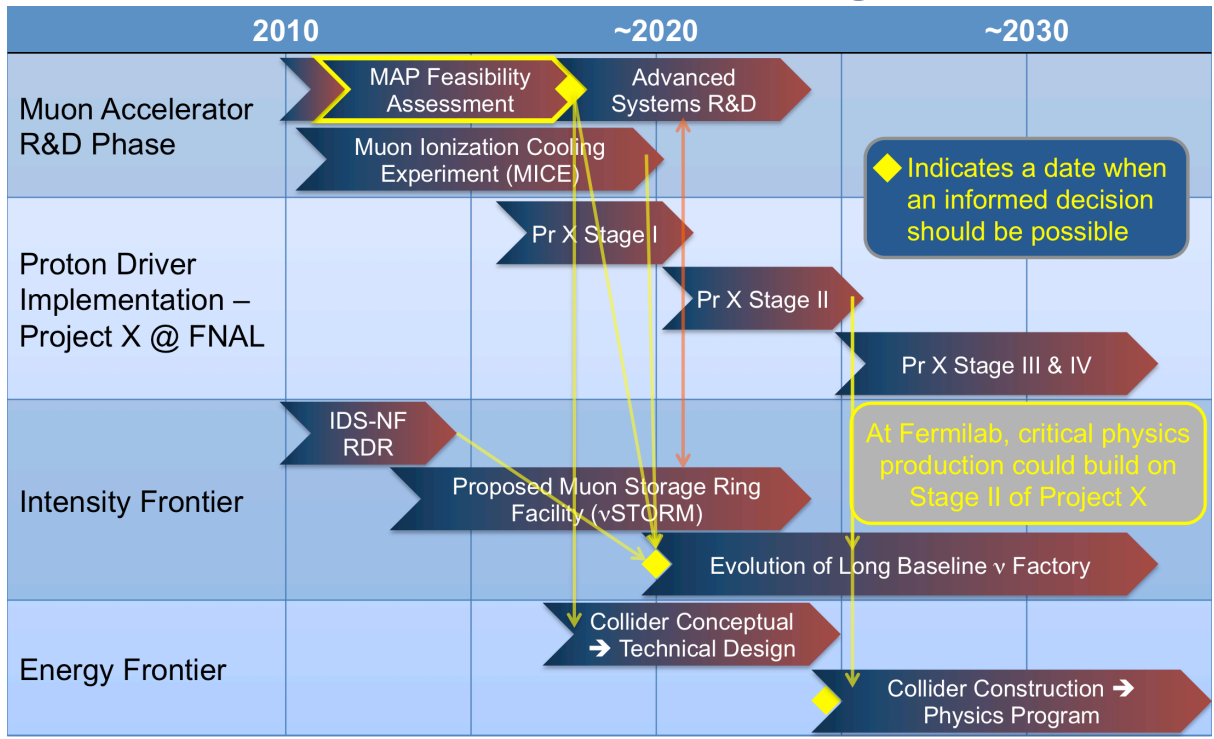

**Figure 1:** Muon accelerator timeline including the MAP Feasibility Assessment period. It is anticipated that decision points for moving forward with a Neutrino Factory project supporting Intensity Frontier physics efforts could be reached by the end of this decade, and a decision point for moving forward with a Muon Collider physics effort supporting a return to the Energy Frontier with a U.S. facility could be reached by the middle of the next decade. These efforts are able to build on Project X Phase II capabilities as soon as they are available. It should also be noted that the development of a short baseline neutrino facility, i.e., nuSTORM, would significantly enhance MAP research capabilities by supporting a program of advanced systems R&D.

---

# U.S. Muon Accelerator Program

## 1.    Landscape of High Energy Physics:

Muon accelerators offer unique potential for the U.S. High Energy Physics community to support a broad and world-leading high energy physics program by enabling a series of staged facilities at both the Intensity and Energy Frontiers.

### 1.1   Intensity Frontier

Neutrino oscillations are irrefutable evidence for physics beyond the Standard Model (SM) of particle physics. The observed properties of the neutrino—the large flavor mixing and the tiny mass—could be consequences of phenomena which occur at energies never seen since the Big Bang. They also could be triggered at energy scales as low as a few keV. Determining the energy scale of the physics responsible for neutrino mass is one of the primary tasks at the Intensity Frontier, which will ultimately require high precision measurements. High precision is necessary since the telltale effects from either a low or high energy scale responsible for neutrino masses and mixing will be very small, either because couplings are very small, as in low-energy models, or the energy scales are very high and thus their effects are strongly suppressed. Neutrino facilities to pursue the study of oscillation phenomena are therefore essential and complementary to high-energy colliders. They are competitive candidates for the next world-class facilities for particle physics.

Within the last 18 months, $\theta_{13}$ has been measured conclusively by reactor antineutrino experiments such as Daya Bay and the angle found to be large, very close to previously established limits. Despite this very large value of $\theta_{13}$, existing beam experiments such as T2K and NOνA will have limited sensitivity to matter–antimatter symmetry (CP) violation and the ordering of neutrino masses (the "mass hierarchy"). With $\theta_{13}$ so large, many alternative mass hierarchy measurement methods have become, at least in principle, feasible. These include the use of
- Atmospheric neutrinos in low-energy upgrades of IceCube;
- Atmospheric neutrinos in ICAL, a 50 kt MINOS-like detector in India;
- Reactor antineutrinos at a distance of about 60 km, the so-called Daya Bay II proposal.

In addition, prospects for uncovering the mass hierarchy by a combination of data from existing experiments, including NOνA in particular, have dramatically increased with the measured value of $\theta_{13}$. As a result, consensus that the mass hierarchy will be determined within the next decade without new beam-based experiments is emerging. At that point, the remaining questions in neutrino oscillation physics will be those of matter-antimatter asymmetries, and whether our current framework of three active neutrinos is complete.

The question whether there are only three neutrinos is underscored by an accumulation of anomalies in short-baseline oscillation experiments: the LSND results, the MiniBooNE event excess, the Reactor Antineutrino Anomaly and the Gallium Anomaly. Each of these seems to point to oscillations with a mass-squared difference of the order of 1 eV$^2$. At the same time, this interpretation is in significant tension with the absence of disappearance effects at the appropriate L/E scale. Such a large mass-squared difference implies the existence of a fourth neutrino, which, due to the LEP results on the invisible Z-decay width, must not couple to the Z boson and hence is not subject to any Standard Model gauge interaction—thus, it is aptly named sterile. A sterile neutrino is the most radical form of physics beyond the Standard Model since it is not part of the framework of gauge symmetries; without gauge symmetries we have no model building tools to constrain the properties of a particle. At the same time it is naive to assume that a sterile neutrino has no other properties beyond its mixing with Standard Model neutrinos. It therefore will be a gateway to a hitherto completely unknown sector of physics.

Both these questions, leptonic CP violation and the completeness of the three-flavor picture, can only by addressed by very high precision measurements of neutrino and antineutrino oscillation probabilities, specifically including channels where the initial and final flavor of neutrino are different. Several neutrino sources have been conceived to reach high sensitivity and to allow the range of measurements necessary to remove all ambiguities in the determination of oscillation parameters. The sensitivity of these facilities is well beyond that of the presently approved neutrino oscillation program. Studies so far have shown that, even for the measured large value of $\theta_{13}$, the Neutrino Factory, an intense high-energy neutrino source based on a stored muon beam, gives the best performance for CP measurements over the entire parameter space. Its time-scale and cost, however, remain important questions. Second-generation superbeam experiments using megawatt proton drivers may be an attractive option in certain scenarios, but eventually the issue of systematics control may limit this technology. It should be noted that once detailed plans are considered, the fiscal and time scales of true superbeams are very large as well.

## 1.2 Energy Frontier

The Standard Model has been a spectacular success. For more than thirty years all new observations have fit naturally into this framework. The recent discovery of a 126 GeV Higgs-like boson at the LHC also appears to be consistent with SM expectations. Furthermore, no evidence of physics beyond the SM (strong dynamics, supersymmetry or extra dimensions) has yet been observed at the ATLAS or CMS experiments. Still, basic questions remain:

- Does this newly discovered boson provide the complete mechanism of electroweak symmetry breaking?
- How do the fermion masses and flavor mixings arise?

Furthermore, the Standard Model is incomplete. It does not explain dark matter; neutrino masses and mixings require new particles or interactions; and the observed baryon asymmetry in the universe requires additional sources of CP violation. From a theoretical viewpoint there are also problems with the SM. It has been argued by G. 't Hooft[1] that the SM is not natural at any energy scale $\mu$ much above the Terascale (1 TeV) because the small dimensionless parameter $\chi^2 = (m_H/\mu)^2$ is not associated with any symmetry in the limit $\chi = 0$. This is the naturalness problem of the SM. If the SM is valid all the way up to the Planck scale $\Lambda_{Pl} \sim 10^{19}$ GeV, then the SM has to be fine-tuned to a precision of one part in $(m_H/\Lambda_{Pl})^{-2}$! In this decade, the physics of the Terascale will be explored at the LHC. Planned experiments studying neutrino oscillations, quark/lepton flavor physics, and rare processes may also provide insight into new physics at the Terascale and beyond.

Discoveries made at the LHC will elucidate the origin of electroweak symmetry breaking. Is that mechanism the SM Higgs scalars or does it involve new physics? New physics might include new gauge bosons, additional fermion generations or fundamental scalars. It might be SUSY or new dynamics or even extra dimensions.

---

# U.S. Muon Accelerator Program

Significant theoretical questions will likely remain even after the full exploitation of the LHC—most notably, the origin of fermion (quark and lepton) masses, mixings and CP violation; the character of dark matter; and detailed questions about spectrum, dynamics, and symmetries of any observed new physics. Thus, it is hard to imagine a scenario in which a multi-TeV lepton collider would not be required in order to fully explore the new physics.

To prepare for the energy frontier in the post-LHC era, research and development are being pursued on a variety of lepton colliders. For the Muon Collider as well as other options a staged approach is envisioned. The first stage is a low energy Higgs Factory at 250–350 GeV for an electron-positron collider [circular (TLEP) or linear (ILC)] or at the s-channel Higgs resonance (~126 GeV) for the Muon Collider. The facility would be planned to be upgradable to a second design capable of higher energies ($E_{cm} < 1$ TeV for ILC or 3 TeV for CLIC) or a multi-TeV Muon Collider. Given the lack of evidence of new physics to date at the LHC, it is prudent to consider the potential energy reach of the various options as an important factor in this choice of future lepton collider. It is possible that scales approaching 10 TeV will be required to fully explore any new physics. In this case, only a Muon Collider could be considered.

A multi-TeV Muon Collider thus provides a very attractive possibility for studying the details of Terascale physics after the LHC. Physics and detector studies are under way to understand the required Muon Collider parameters (in particular luminosity and energy) and to map out, as a function of these parameters, the associated physics potential. The physics studies will set benchmarks for various new physics scenarios (e.g., SUSY, Extra Dimensions, New Strong Dynamics) as well as Standard Model processes.

## 1.3    The Beauty and Challenges of Muon-based Facilities

Muon-based facilities offer the unique potential, among the accelerator concepts being discussed in the Community Summer Study process, to provide the next generation of capabilities and world-leading experimental support spanning physics at both the Intensity and Energy Frontiers. Building on the foundation of Project X at FNAL, muon accelerators can provide that next step with a high-intensity and precise source of neutrinos to support a world-leading research program in neutrino physics. Furthermore, the infrastructure developed to support such an Intensity Frontier research program can also enable the return of the U.S. high energy physics program to the Energy Frontier. This capability would be provided in a subsequent stage of the facility that would support one or more Muon Colliders, which could operate at center-of-mass energies from the Higgs resonance at 126 GeV up to the multi-TeV scale, if and when required for studies beyond the Standard Model.

Pending the needed technology feasibility demonstrations, Muon Colliders would constitute the ideal facilities to explore the multi-TeV colliding beam energy range since
- They profit from multi-turn collisions and multiple interaction points as circular colliders but without energy limitation by emission of synchrotron radiation;
- They do not suffer from beamstrahlung as do linear colliders.

Consequently, they present great potential for
- Large luminosity integrated over several detectors for support of a broad physics community;
- An attractive energy spectrum with small momentum spread at the collision point due to the absence of beamstrahlung;
- Limited power consumption due to multi-turn collisions;

- Affordable cost due to the limited physical size of the facilities.

An ensemble of facilities possibly built in stages is made possible by the strong synergies between Neutrino Factories and Muon Colliders, both of which require a high power proton source and target for muon generation followed by similar front-end and ionization cooling channels. It is especially attractive at FNAL taking advantage of the proton driver potential of Project X and the ability to deploy an optimized detector at SURF.

As developed in the following sections, these muon facilities rely on a number of systems with conventional technologies whose required operating parameters exceed the present state of the art as well as novel technologies unique to muon colliders. An R&D program to evaluate the feasibility of these technologies is being actively pursued within the framework of the U.S. Muon Accelerator Program (MAP). The critical challenges include:

- A high-power proton linac and target station (up to 4 MW) although full power capability is not required for initial Neutrino or Higgs Factory operation;
- A 15–20 T capture solenoid;
- RF accelerating gradient in low frequency (325–975 MHz) structures immersed in high magnetic field as required for the front end and ionization cooling sections;
- Ionization cooling by 6 orders of magnitude (2 in each transverse plane and 2 in longitudinal plane);
- Very high field (> 30 T) solenoids utilizing high temperature superconducting (HTS) coils (only required for the multi-TeV collider final cooling section);
- Recirculating linacs (RLA) and rapid cycling synchrotron (RCS) or fixed-field alternating-gradient (FFAG) rings for fast beam acceleration;
- A collider ring design and machine–detector interface (MDI) including absorbers for the decay products of the muon beams;
- Detector operation in a unique background environment caused by the muon decays around the ring.

# 2   A Staged Muon-Based Facility Program

## 2.1   Rationale for a Staged Approach

The feasibility of the technologies required for Neutrino Factories and/or Muon Colliders must be validated before a facility based upon these could be proposed. Such validation is usually made in dedicated test facilities, which are specially designed to address the major issues. Although very convenient, these test facilities are rather expensive to build and to operate over several years. They are therefore difficult to justify and fund, given especially that they are usually useful only for technology development rather than for physics.

An alternative approach is proposed here. It consists of a series of facilities built in stages, where each stage offers

- Unique physics capabilities such that the corresponding facility obtains support and can be funded.
- In parallel with the physics program, integration of an R&D platform using each stage as a source of particles to develop, test with beam and validate a new technology that will be necessary for the following stages.
- The system based on the novel technology, once proved to work, and even if not necessary for the present stage, could be used to improve its performance.

# U.S. Muon Accelerator Program

- Operation of the novel technology in a realistic environment—extremely useful not only to validate the novel technology itself, but also to acquire operational experience before using it in the following stage.
- Construction of each stage as an add-on to the previous stages, extensively reusing the equipment and systems already installed, such that the additional budget of each stage remains affordable.

The staging plan we discuss builds on, and takes advantage of, existing or proposed facilities at FNAL, thus maximizing the synergies between the existing FNAL program and the foreseen MAP program, specifically:

- Existing tunnels and other conventional facilities;
- Project X at Fermilab as the MW-class proton driver for muon generation;
- SURF as developed for the LBNE detector, which could then house the detector for a long-baseline Neutrino Factory (which could initially be the LBNE detector itself).

The plan consists of a series of facilities with increasing complexity, each with performance characteristics providing unique physics reach:

- nuSTORM:  a short-baseline Neutrino Factory-like ring enabling a definitive search for sterile neutrinos, as well as neutrino cross-section measurements that will ultimately be required for precision measurements at any long-baseline experiment.
- NuMAX:  an initial long-baseline Neutrino Factory, optimized for a detector at SURF, affording a precise and well-characterized neutrino source that exceeds the capabilities of conventional superbeam technology.
- NuMAX+:  a full-intensity Neutrino Factory, upgraded from NuMAX, as the ultimate source to enable precision CP-violation measurements in the neutrino sector.
- Higgs Factory:  a collider whose baseline configurations are capable of providing between 3500 (during startup operations) and 13,500 Higgs events per year ($10^7$ sec) with exquisite energy resolution.
- Multi-TeV Collider:  if warranted by LHC results, a multi-TeV Muon Collider likely offers the best performance and least cost for any lepton collider operating in the multi-TeV regime.

Each stage is described below in terms of

- Physics interest;
- Facility and detector;
- The required R&D;
- The possible technology validation for the following stage.

Their main parameters and performance are described in Tables 1 and 2 of the Executive Summary.  A complex integrating all of the above facilities in a staged approach integrates well with Project X on the FNAL site as shown in Figure 2.

# U.S. Muon Accelerator Program

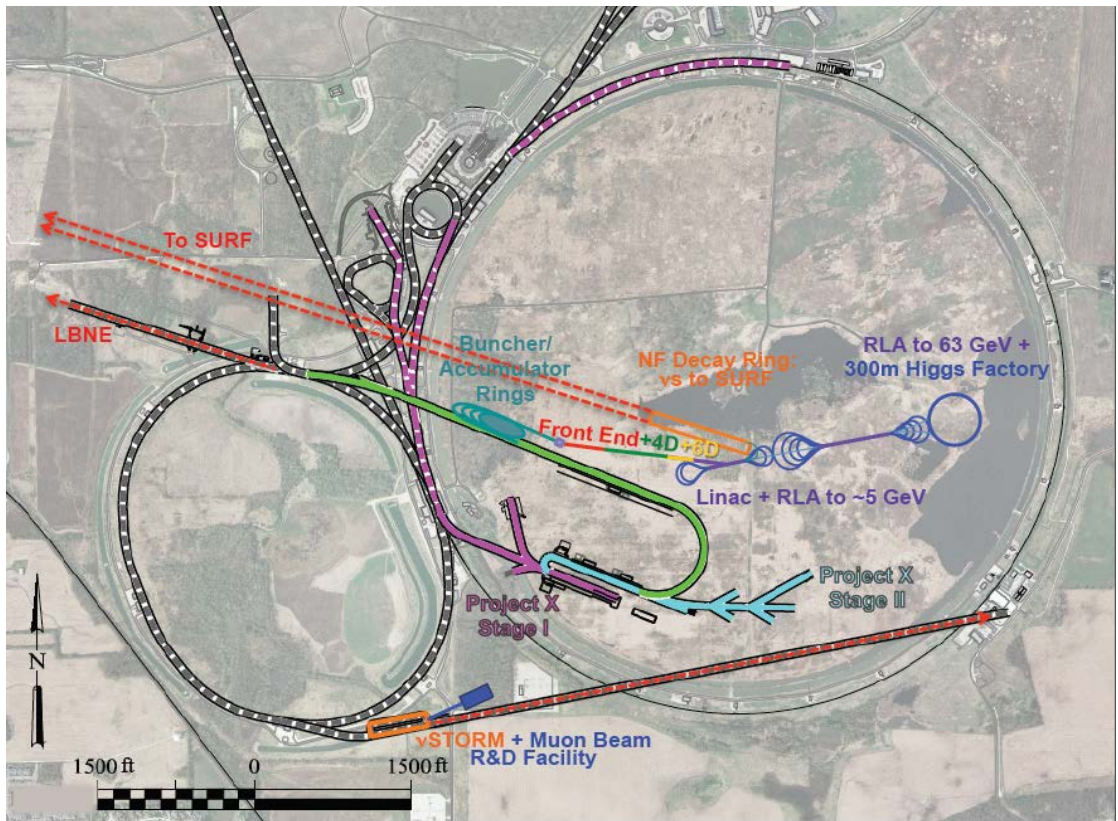

**Figure 2:** Footprint of Neutrino Factory and Higgs Factory Muon Collider facilities on the FNAL site.

## 2.2  nuSTORM

### 2.2.1  Overview

The idea of using a muon storage ring to produce a high-energy ($\approx 50$ GeV) neutrino beam for experiments was first discussed by Koshkarev[2] in 1974. Neuffer first produced a detailed description of a muon storage ring for neutrino oscillation experiments[3] in 1980. In his paper, Neuffer studied muon decay rings with $E_\mu$ of 8, 4.5 and 1.5 GeV. His 4.5 GeV design achieved approximately $6 \times 10^9$ useful neutrinos per $3 \times 10^{13}$ protons on target. The facility we describe here (nuSTORM), essentially the same as that proposed in 1980, will utilize a 3–4 GeV/c muon storage ring to study eV-scale oscillation physics and, in addition, could add significantly to our understanding of $\nu_e$ and $\nu_\mu$ cross sections. In particular, it can

- Serve a first-rate neutrino-physics program, encompassing
  - ➢ Exquisitely sensitive searches for sterile neutrinos in both appearance and disappearance modes;
  - ➢ Detailed and precise studies of electron- and muon-neutrino–nucleus scattering over the energy range required by the future long- and short-baseline neutrino oscillation program; and

---

[2] Proposal for a Decay Ring to Produce Intense Secondary Particle Beams at the SPS, CERN/ISR-DI/74-62, 1974.

[3] D. Neuffer, "Design Considerations for a Muon Storage Ring," Telemark Conference on Neutrino Mass, V. Barger and D. Cline (eds.), Telemark, WI, 1980.

# U.S. Muon Accelerator Program

- Provide the technology test-bed required to carry out the R&D critical for the implementation of the next step in a muon-accelerator based particle-physics program.

The facility can be viewed as the simplest implementation of the Neutrino Factory concept described by Geer[4]. In our case, 120 GeV/c protons are used to produce pions off of a conventional solid target. The pions are collected with a horn and are then transported to, and injected into, a storage ring. Pions that decay in the first straight of the ring can yield muons that are captured in the ring. The circulating muons then decay into electrons and neutrinos. We are starting with a storage ring design that is optimized for 3.8 GeV/c muon momentum. This momentum was selected to maximize the physics reach for both oscillation and cross section physics. Figure 3 shows a schematic diagram of the facility.

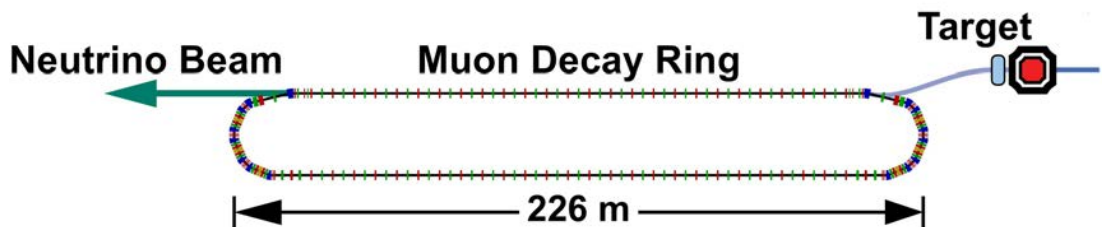

**Figure 3:** Schematic diagram of the nuSTORM layout.

In nuSTORM, the neutrinos are produced by the purely leptonic, and therefore well understood, decay of muons, and thus the neutrino flux can be known with very high, sub-percent, precision. The signals are wrong-sign muons that can be identified quite easily in a magnetized iron detector. The precise knowledge of the neutrino flux and the expected very low backgrounds for the wrong-sign muon search allow one to reduce systematic effects to a negligible level, hence permitting a precise measurement of the new physics that may be behind the short-baseline anomalies. The possible exclusion regions for sterile-neutrino oscillation parameters obtained from 5 years of nuSTORM running are shown in Figure 4.

Muon decay yields a neutrino beam of precisely known flavor content and energy. In addition, if the circulating muon flux in the ring is measured accurately (with beam-current transformers, for example), then the neutrino beam flux is also accurately known. Near and far detectors are placed along the line of one of the straight sections of the racetrack decay ring. The near detector can be placed 20–50 meters from the end of the straight. A near detector for disappearance measurements will be identical to the far detector, but only about one-tenth the fiducial mass. It will require a muon catcher, however. Additional purpose-specific near detectors can also be located in the near hall and will measure neutrino–nucleon cross sections. nuSTORM can provide the first precision measurements of $\nu_e$ and $\nu_e$-bar cross sections—important for future long-baseline experiments. A far detector at approximately 2000 m will study neutrino oscillation physics and be capable of performing searches in both appearance and disappearance channels. The experiment will take advantage of the "golden channel" of oscillation appearance, $\nu_e \to \nu_\mu$, where the resulting final state has a "wrong-sign" muon, of opposite sign as those from interactions of the $\nu_\mu$-bar in the beam (e.g., in the case of $\mu^+$ stored in the ring, this would mean the observation of an event with a $\mu^-$). The detector will thus need to be magnetized in order to identify the wrong-sign muon appearance channel, as is the case for the current baseline Neutrino

---

Factory detector[5]. A number of possibilities for the far detector exist. However, a magnetized iron detector ("MIND") similar to that used in MINOS is likely to be the most straightforward approach. For the purposes of nuSTORM oscillation physics, a detector inspired by MINOS, but with thinner plates and much larger excitation current (larger B field), is assumed ("SuperBIND").

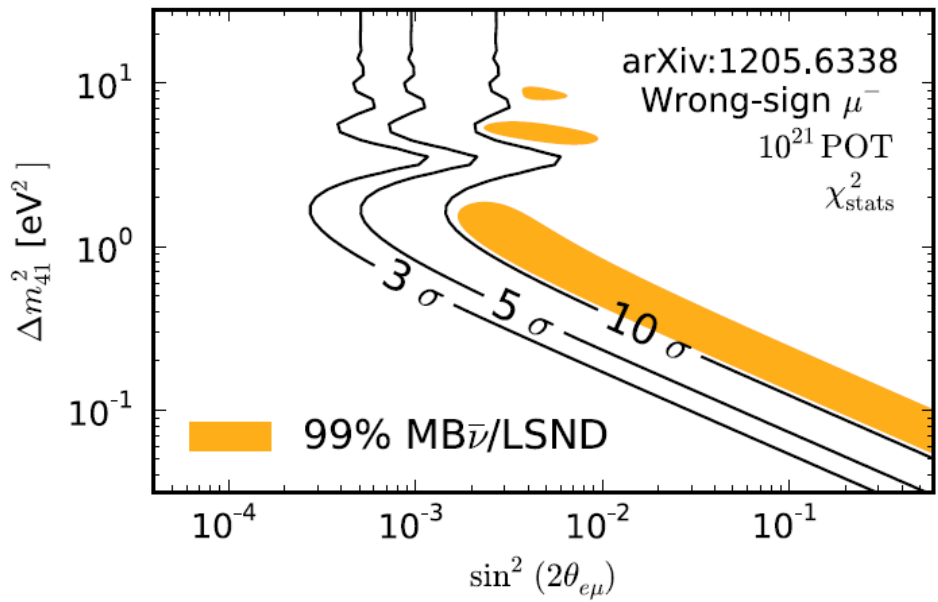

**Figure 4:** Exclusion limits from a five-year run of nuSTORM (statistical only). The orange/shaded areas show the combined 99%-confidence-level-allowed regions from MiniBooNE and LSND.

### 2.2.2 The detector

The SuperBIND detector concept for nuSTORM oscillation physics is shown schematically in Figure 5. The iron plates are disks with an overall diameter of 6 m and thickness of 1.5 cm. (Detector performance for 1 and 2 cm thick plates has also been simulated.) We envision that no R&D on the iron plates will be needed. Final specification of the plate structure will be determined once a plate fabricator is chosen.

As mentioned above, SuperBIND will have a toroidal magnetic field like that of MINOS. For excitation, however, we plan to use the concept of the superconducting transmission line (STL) developed for the Design Study for a Staged Very Large Hadron Collider[6]. Minimization of the muon charge misidentification rate requires the highest field possible in the iron plates. SuperBIND thus requires a much larger excitation current per turn than that of the MINOS near detector (40 kA-turns). A configuration with 8 turns (operating at 30kA) of the STL with a 20 cm hole has been simulated. Figure 6 shows the results of a 2D finite-element magnetic-field analysis utilizing the plate geometry of Figure 5.

# U.S. Muon Accelerator Program

Particle detection using extruded scintillator and optical fibers is a mature technology and has been used in many experiments including MINOS, Scibar, INGRID, P0D, ECAL and the Double-Chooz cosmic-ray veto detectors. Our initial concept for the readout planes is to have both an x and a y view following each plate. Given the rapid development in recent years of solid-state photodetectors based on Geiger mode operation of silicon avalanche photodiodes, this technology has been chosen for SuperBIND.

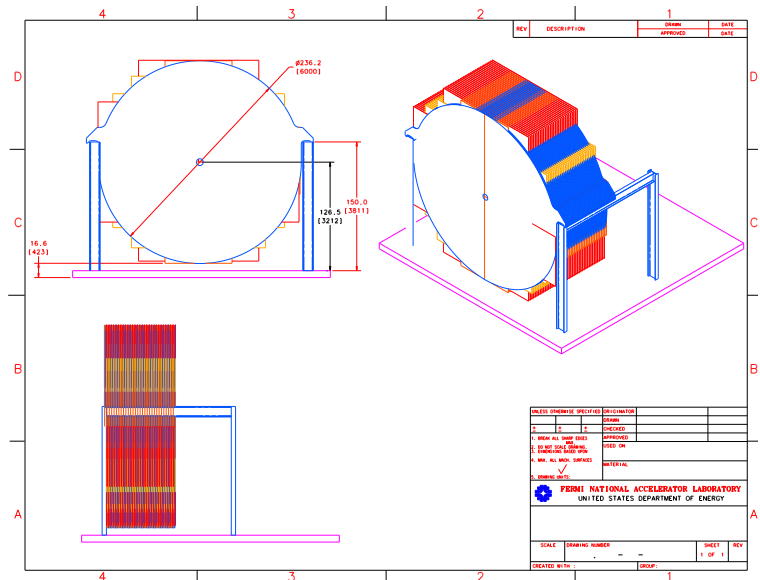

**Figure 5:** Schematic diagram of SuperBIND.

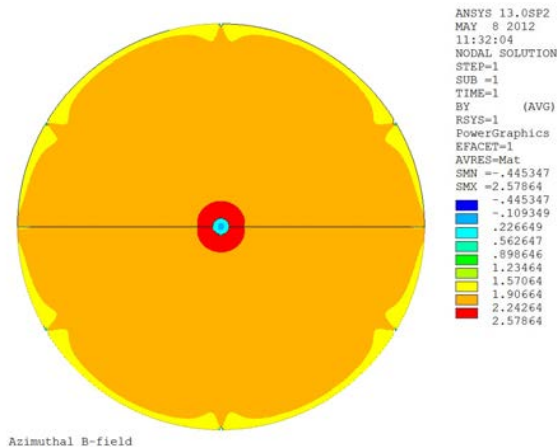

**Figure 6:** Field map in the SuperBIND detector assuming a 20 cm diameter hole and the CMS steel B-H curve, with an excitation current of 240 kA-turns.

### 2.2.3   The facility

The basic concept for the facility is presented in Figure 3 and its main parameters summarized in Table **3**. A high-intensity proton source places beam on a target, producing a broad spectrum of

# U.S. Muon Accelerator Program

secondary pions. Forward pions are focused by a horn into a transport channel. Pions decay within the first straight of the decay ring and a fraction of the resulting muons are stored in the ring. Muon decay within the straight sections will produce neutrino beams of known flux and flavor. For the implementation described here, we choose a 3.8 GeV/c storage ring to obtain the desired spectrum of $\approx$ 2–3 GeV neutrinos. This means that pions must be captured at a momentum of approximately 5 GeV/c.

**Table 3:** nuSTORM parameters

| Parameter | Unit | Value |
|---|---|---|
| Muon momentum | GeV/c | 3.8 |
| Momentum acceptance | % | 10 |
| Proton momentum | GeV/c | 120 |
| Power on target | kW | 100–150 |
| Muons | per pulse | $5 \times 10^{10}$ |
| Ring circumference | m | 470 |
| Detector mass (Far) | kt | 1.3 |
| Detector mass (Near) | kt | 0.2 |

The number of pions produced by 60–120 GeV/c protons with various targets has been simulated using the MARS code[7] giving the pion rate, per proton on target, in a forward cone of 120 mrad. A target optimization based on a conservative estimate for the decay-ring acceptance of 2 mm·rad was then done which indicated that a yield of approximately 0.10 pions per proton on target can be collected into a ±10% momentum acceptance off of medium/heavy targets assuming 80% capture efficiency.

An obvious goal for the facility is to collect as many pions as possible (within the limits of available beam power), inject them into the decay ring and capture as many muons as possible from the $\pi \rightarrow \mu$ decays. With pion decay within the ring, non-Liouvillean "stochastic injection" is possible. In stochastic injection, the ~ 5 GeV/c pion beam is transported from the target into the storage ring and dispersion-matched into a long straight section. (Circulating and injection orbits are separated by momentum.) Decays within that straight section provide muons that are within the ~ 3.8 GeV/c ring momentum acceptance (see Figure 7). For 5.0 GeV/c pions, the decay length is ~ 280 m, thus ~ 52% decay within the 210 m decay ring straight.

The baseline for the muon decay ring is a FODO racetrack, although our Japanese collaborators are also investigating an FFAG racetrack. The FODO ring (Figure 8) uses both normal and superconducting magnets. A FODO lattice, using only normal-conducting magnets (B ≤ 2 T), is also being developed, giving arcs that are twice as long (~50 m), but the straight sections would be similar.

The design goal for the ring was to maximize both the transverse and momentum acceptances, while maintaining reasonable physical apertures for the magnets in order to keep the cost down. This was accomplished by employing strongly focusing optics in the arcs (90° phase advance per FODO cell), featuring small beta functions (approximately 3 m average) and low dispersion (approximately 0.8 m average).

---

[7] N.V. Mokhov, "The Mars Code System User's Guide," Fermilab-FN-628, 1995.

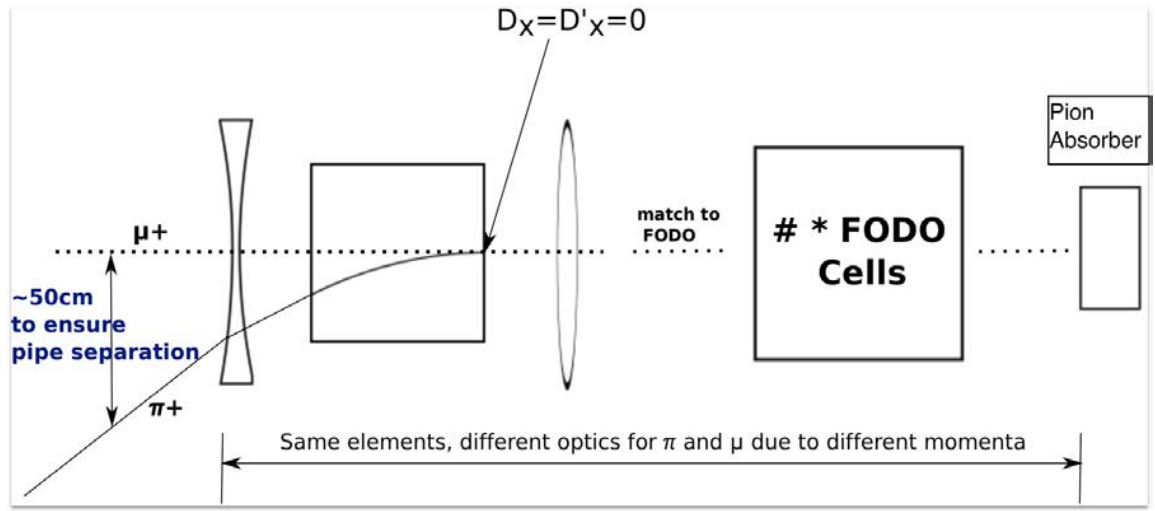

**Figure 7:** Stochastic injection scheme

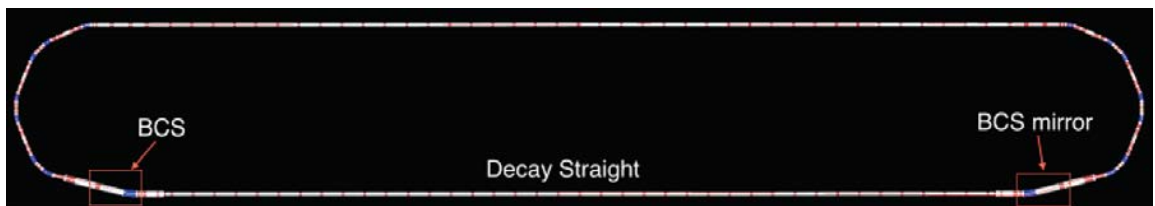

**Figure 8:** FODO decay ring schematic diagram

### 2.2.4    Required R&D

No R&D is required for nuSTORM.  Some magnet prototyping work will be required, however.  Most of the facility's components (primary proton beam line, target station, civil construction, detector) have been done before and existing technology is entirely suitable.

### 2.2.5    Siting at Fermilab

nuSTORM will use the Main Injector (MI) abort line to extract protons from the MI through an existing beam pipe in the MI abort absorber to a new target station to the southeast.  The transport line and decay ring are positioned on the Fermilab site east of Kautz Road.  The near detector hall is located 20 m from the end of the production straight and the far detector will be located in the existing D0 Assembly Building (DAB).  A photo of the Fermilab site with the nuSTORM facilities superimposed is shown in Figure 9.

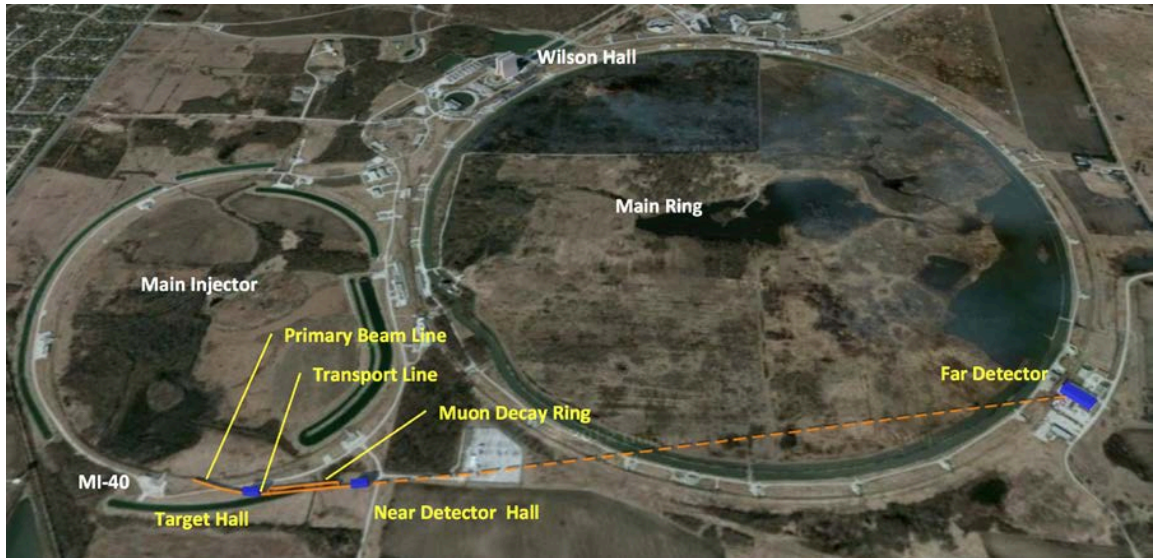

**Figure 9:** nuSTORM facilities superimposed on the Fermilab site

### 2.2.6   Technology validation for following phases

Advanced R&D for the high-intensity 6D ionization cooling channel required for a Muon Collider could be pursued using the nuSTORM facility, which provides a muon source with significant intensity ($\approx 10^{10}$ μ/pulse in the 100–300 MeV/c momentum range). This beam can be produced simultaneously with the neutrino physics program at little additional cost. This is possible because nuSTORM requires an absorber to absorb pions remaining (about 60% of those injected into the ring) after the first straight (see Figure 10). Pions in the momentum range 5 GeV/c ±10% are extracted to the absorber. There are also many muons in the same momentum window (forward decays) that will be extracted along with the pions. The absorber will act as a degrader for these muons, producing the desired low-energy muon beam. Figure 11 shows the muon momentum distribution after an absorber consisting of 3480 mm of Fe. In addition, nuSTORM will present the opportunity to design, build and test decay ring instrumentation (BCT, momentum spectrometer, polarimeter) to measure and characterize the circulating muon flux.

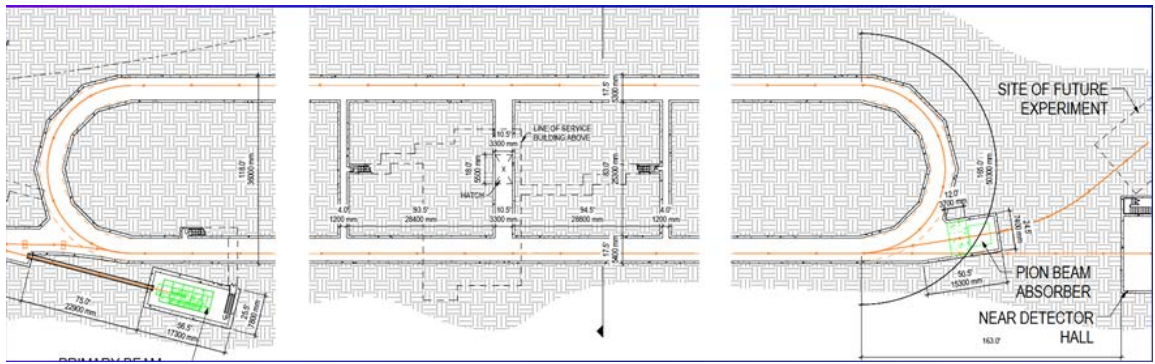

**Figure 10:** nuSTORM storage ring. About 50% of pions decay in the straight section. Pion absorber serves as a degrader to produce muons in the desired momentum range.

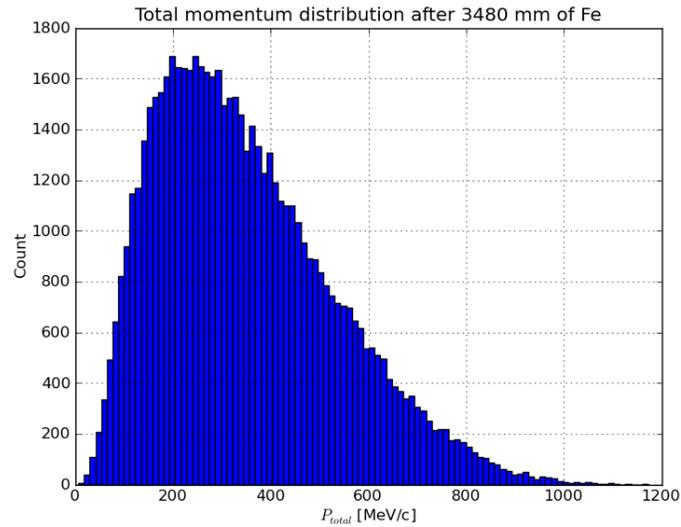

**Figure 11:** Muon beam momentum distribution at the exit of the degrader. $O(10^{10})$ muons per pulse should be available.

Two key 6D cooling channel designs currently under detailed study can be tested at the nuSTORM facility without affecting the main neutrino activities: the Guggenheim and the Helical Cooling Channel (HCC); see layouts in Figure 12. Once the bench test for one of these channels is carried out with no beam, a section of cooling channel long enough for appreciable 6D cooling could be used at the nuSTORM facility for a demonstration with beam. The nuSTORM facility could also house the equipment and infrastructure required for the initial bench test.

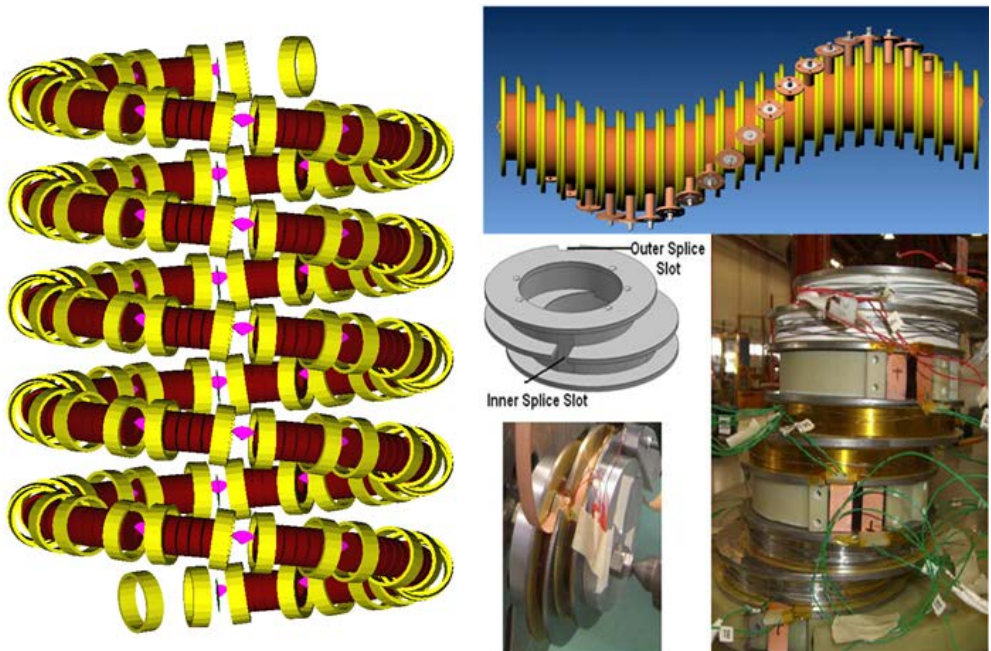

**Figure 12:** Two six-dimensional cooling channel designs. Left: initial stages of the Guggenheim channel; top-right: conceptual drawing of the Helical Cooling Channel; bottom-right: HCC test coil assembly.

# U.S. Muon Accelerator Program

## 2.3   Neutrino Factory

### 2.3.1   Physics specific to muon based Neutrino Factory

The Neutrino Factory concept is attractive since it provides very high intensity neutrino and antineutrino beams which are exact CP conjugates. The flavor content and energy spectrum as well as the total flux can be determined to better than 1%, which, combined with the great flexibility in neutrino energy, makes a Neutrino Factory the ideal source for precision neutrino physics. Moreover, the beam contains equal numbers of muon and electron flavors and therefore, it is possible to directly measure the relevant cross sections, including nuclear effects, in the near detector. As a result it is widely recognized that the Neutrino Factory is the only concept that will allow an accuracy in the determination of leptonic mixing parameters that can compete with that in the quark sector.

Neutrino Factories were originally designed to cover the smallest possible values of $\theta_{13}$, but in response to the measurement of large $\theta_{13}$, the Neutrino Factory design was reoptimized to a stored muon energy of 10 GeV and a single baseline of 2000 km using a 100 kt magnetized iron detector. It is possible to further reduce the energy to around 5 GeV and concomitantly the baseline to 1300 km without an overall loss in performance if one changes the detector technology; possible choices include a magnetized liquid argon or fully active plastic-scintillator detector to improve efficiency around 1–2 GeV. Once one of these technology choices is shown to be feasible, there is no strong physics-performance reason to favor the 10 GeV over the 5 GeV option, or vice versa. The low-energy option is attractive due to its synergies with planned super-beams such as LBNE and because the detector technology would allow a comprehensive physics program in atmospheric neutrinos, proton decay and supernova detection. For the low-energy option detailed studies of intensity staging have been carried out which indicate that even at 1/20[th] of the full-scale beam intensity and starting with a 10 kt detector, significant physics gains beyond the initial phases of a pion-decay based experiment, such as LBNE, can be realized. At full beam intensity and with a detector mass in the range of 10–30 kt, a 5 GeV Neutrino Factory offers the best performance of any conceived neutrino oscillation experiment (Figures 13 and 14).

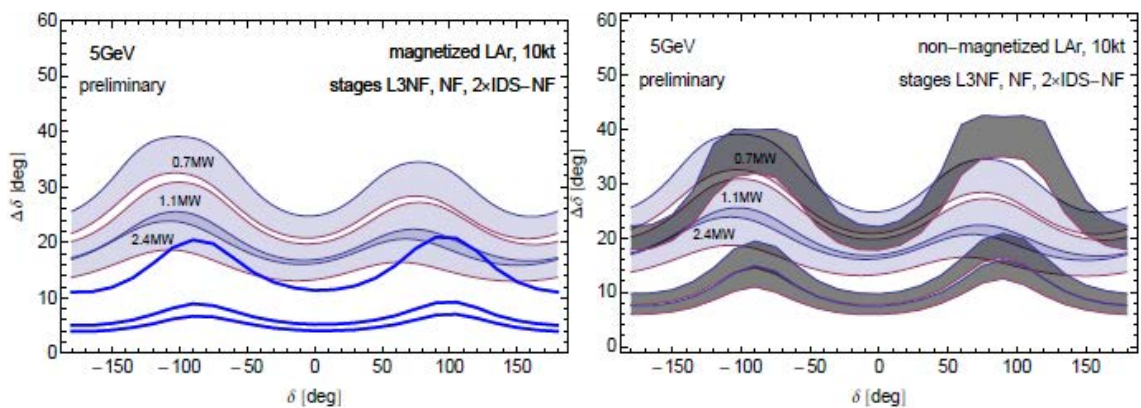

**Figure 13:** Accuracy on the CP phase vs. the true value of the CP phase at 1σ confidence level. Light-blue bands depict the accuracy expected from LBNE using the various beams Project X can deliver. In the left panel, the thick blue curves represent what a Neutrino Factory beam can do using a magnetized LAr detector. In the right panel, the gray bands illustrate the accuracy of a Neutrino Factory using a non-magnetized detector (for Neutrino Factory beam intensities see Table 1 of the Executive Summary).

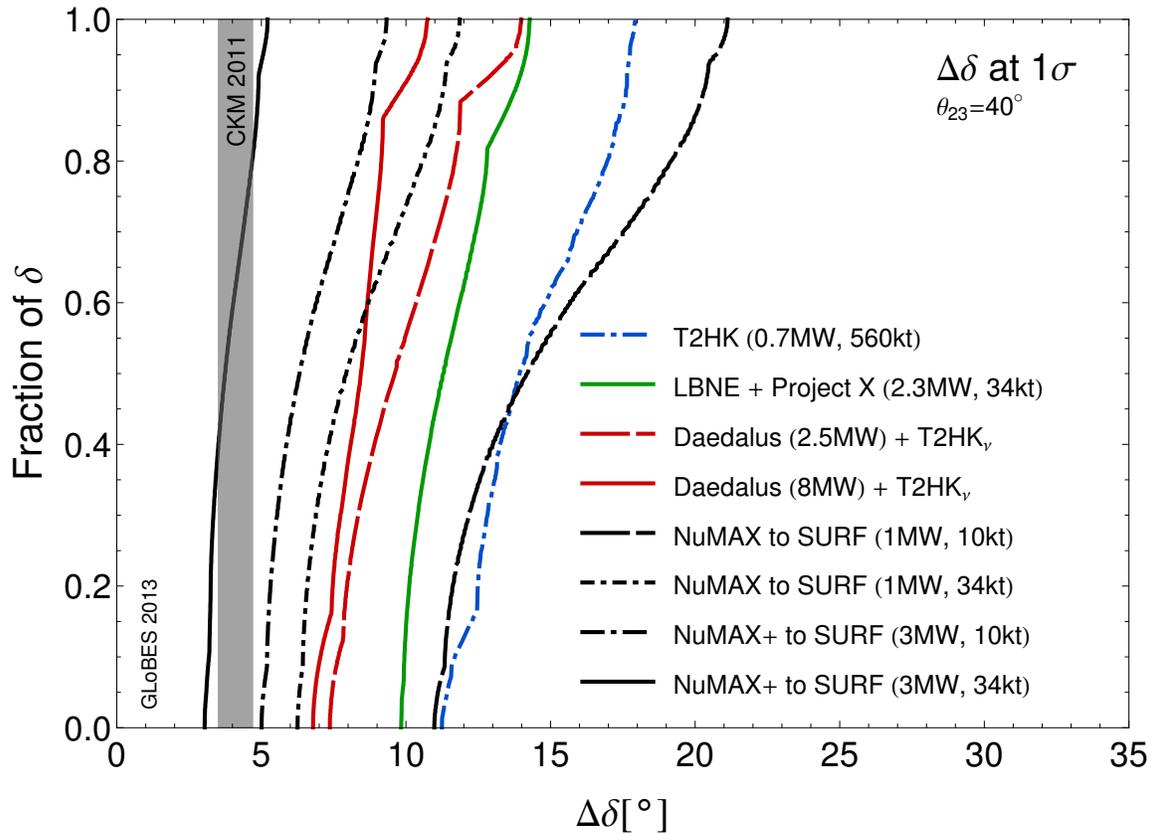

**Figure 14:** Comparison of the sensitivity to CP violating for various experimental schemes including LBNE with Project X, T2HK, Daedelus and the various possible stages of NuMAX, where the NuMAX sensitivities assume a magnetized detector.

To pursue the study of short-baseline neutrino oscillations, several proposals exist, both at FNAL and CERN, to use pion decay-in-flight beams, as MiniBooNE did; the crucial difference with respect to MiniBooNE would be the use of a near detector and the potential use of LAr TPCs instead of scintillator detectors. While these new proposals would constitute a significant step beyond what MiniBooNE has achieved, especially in terms of systematics control, it remains to be proven that a beam which has a 1% level contamination of $\nu_e$ can be used to perform a high-precision study of a sub-percent $\nu_e$ appearance effect. In particular, it should be pointed out that many of these proposals involve near and far detectors of very different sizes and/or geometrical acceptance, and thus cancellations of systematics will be far from perfect. Therefore, it is not obvious that these experiments can take full advantage of the beam intensities Project X will deliver.

### 2.3.2 The detector in a phased approach

With a baseline of 1300 km the relevant neutrino energies for oscillation measurements (dictated by $\Delta m_{32}^2$) lie in the 1–2 GeV range. The MIND technology preferred for the International Design Study for the Neutrino Factory (IDS-NF) starts to become inefficient at these low energies and it is anticipated that a change of detector technology will be needed. Two candidates suggest themselves at this point in time: magnetized, fully active, plastic scintillator and magnetized liquid argon TPCs. Since LBNE has chosen a liquid argon (LAr) TPC (Figure 15) as its far-detector technology, a staged approach to a Neutrino Factory using a magnetized liquid argon detector seems the way to go, with possibly 10 kt fiducial mass (twice as much for the whole detector) at NuMAX upgradable to 30 kt at NuMAX+.

There is considerable liquid argon TPC R&D taking place worldwide with the primary goal of providing input to the detailed design of the LBNE far detector(s). There have been some R&D efforts in Europe toward a magnetized LAr TPC, but considerable R&D remains to be done. Pending that R&D, it is not yet clear whether a non-magnetized LAr TPC for LBNE could be economically retrofitted with a magnetic field or whether an entirely new detector would need to be built.

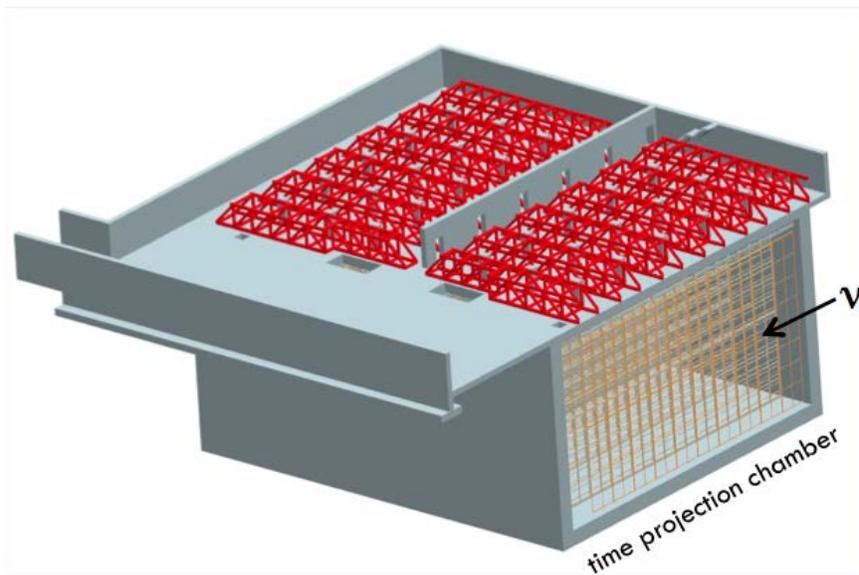

**Figure 15:** The 10 kt liquid argon LBNE detector

### 2.3.3    The facilities in a phased approach

#### 2.3.3.1    Introduction

Here we describe facilities required for a staged approach to a Neutrino Factory, which could eventually be reused for future Higgs Factory and multi-TeV Muon Colliders. The proposed staging scenario envisions first a lower-energy (5 GeV), lower-intensity Neutrino Factory (NuMAX), upgradable to full intensity (NuMAX+), with parameters listed in Table 1.

The Neutrino Factory uses a high-energy proton beam to produce charged pions. The majority of the produced pions have momenta of a few hundred MeV/c, with a large momentum spread, and transverse momentum components that are comparable to their longitudinal momentum. Hence, the daughter muons are produced within a large longitudinal and transverse phase-space. This initial muon population must be confined transversely, captured longitudinally, and have its phase-space manipulated to fit within the acceptance of an accelerator. These beam manipulations must be done quickly, before the muons decay ($\tau_0 = 2.2$ μs). Finally, muons are stored in the decay ring to produce neutrino beams in the ring's straight sections. The figure of merit describing performance of the various stages is the neutrino flux generated by decaying muons in the storage ring straights. Assuming a standard $10^7$ operating seconds/year, the

# U.S. Muon Accelerator Program

projected muon fluxes, as summarized in Table 1, are 2 ×10²⁰ μ± per year (NuMAX) and 1.2 × 10²¹ μ± per year (NuMAX+).

### 2.3.3.2    Components

The functional elements of a Neutrino Factory, illustrated schematically in Figure 16, are as follows:

- A proton source producing a high-power multi-GeV bunched proton beam.

- A pion production target that operates within a high-field solenoid. The solenoid confines the pions radially, guiding them into a decay channel.

- A solenoid decay channel.

- A system of RF cavities that captures the muons longitudinally into a bunch train, and then applies a time-dependent acceleration that increases the energy of the slower (low-energy) bunches and decreases the energy of the faster (high-energy) bunches.

- A cooling channel that uses ionization cooling to reduce the transverse phase space occupied by the beam, so that it fits within the acceptance of the first acceleration stage.

- An acceleration scheme that accelerates the muons to 5 GeV.

- A 5 GeV "racetrack" storage ring with long straight sections.

For the sake of an early start, NuMAX is similar to NuMAX+ except for the proton driver, with a reduced beam power of 1 MW instead of 3 MW, and no muon cooling. Its performance is therefore reduced by about a factor of six: 3 due to the reduced proton beam power and 2 due to the lack of cooling.

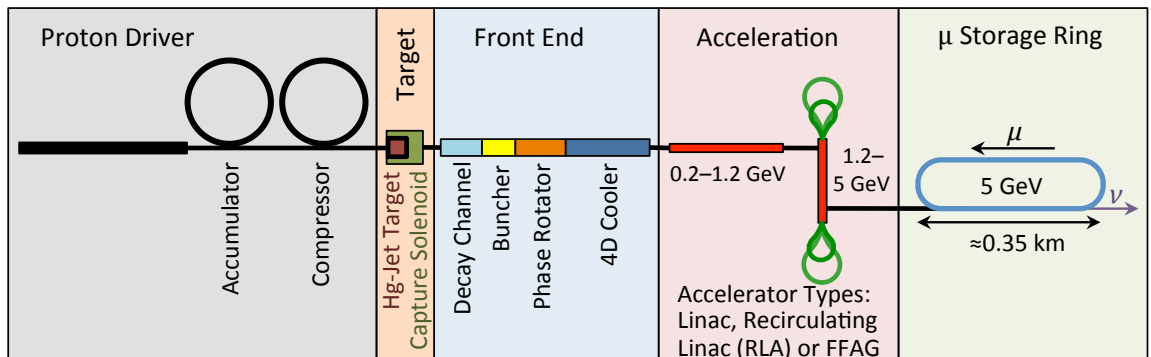

**Figure 16:** Functional elements of a 5 GeV Neutrino Factory

### 2.3.3.3    Implementation on the Fermilab site

Here we discuss facility specifics based on Fermilab's infrastructure and the various stages of Project X. The facilities will support NuMAX and its upgrade to a full-intensity NF at 5 GeV. A schematic view of the facility layout on the Fermilab site is given in Figure 2. The above scheme and its components are described below.

# U.S. Muon Accelerator Program

2.3.3.3.1   Proton Driver

The primary requirement for an NF Proton Driver is the number of useful muons produced at the end of the decay channel, which, to good approximation, is proportional to the primary proton beam power, and (within the 5–15 GeV range) only weakly dependent on the proton beam energy. Studies have shown that proton beam power in the 1–4 MW range is needed[8,9]. In addition to the beam-power requirement, short proton bunches, 2±1 ns (rms), are required.

Thus, the 3 GeV/1 MW proton beam provided by the Project X Stage IIa linac is suitable for NuMAX. Nevertheless, higher beam power is highly desirable. Upgrading to the full 3 MW capability of Project X Stage IIb will improve the neutrino flux by a factor of 3. Further upgrades later to the 4 MW beam power at 8 GeV provided by Project X Stage IV would increase the neutrino flux by about another factor of 2.

The proton beam must be bunched to form a bunch structure suitable for a NF. A bunching scheme based on two rings can be implemented as shown in Figure 16. The first storage ring (the Accumulator) accumulates, via charge stripping of the H⁻ beam. The incoming beam from the linac is chopped to allow clean injection into pre-existing RF buckets. Painting will be necessary in 4D transverse phase space, and possibly also in longitudinal phase-space, in order to control space-charge forces. The second storage ring (the Compressor) accepts two to four bunches from the Accumulator and then performs a 90º bunch rotation in longitudinal phase space, shortening the bunches at the limit of space-charge tune shift just before extraction. The ring must have a large momentum acceptance to allow for the beam momentum spread (a few %) during bunch rotation. The short bunches are extracted from the Compressor into separate ("trombone") transport lines of differing lengths so that they arrive on the target at the same time.

2.3.3.3.2   Target and Decay Channel

Results from the MERIT (Mercury Intense Target) Experiment[10] have provided a proof-of-principle demonstration for a free Hg-jet target technology that could survive beam power up to ∼ 8 MW, as contemplated in NF scenarios. The target, pion-collection, and pion-decay channel for high-energy Neutrino Factories have been extensively studied. They involve short (1 to 3 ns rms) pulses of protons focused onto a liquid Hg-jet target immersed in a high-field (20 T) solenoid. The same designs can be used for a low-energy NF using a 3–8 GeV proton source, and therefore the design from the IDS-NF[11] can be adopted.

The initial proton bunch is relatively short, and as the secondary pions drift from the target they spread longitudinally. Hence, downstream of the target, the pions and their daughter muons develop a position–energy correlation in the (RF-free) decay channel. In the IDS-NF baseline design, the drift length is chosen as 56.4 m, and at the end of the decay channel there are about 0.2 muons of each sign per incident 8 GeV proton[12].

The early NuMAX phase based on a proton driver of 1 MW at 3 GeV could start with a more conventional target or could take advantage of the experience gained on the Hg target operated by SNS at similar power for several years. The evolution from NuMAX to NuMAX+ based on a power increase of the proton beam on target to the 3 to 4 MW range will require a major target upgrade.

### 2.3.3.3.3    Bunching and Phase Rotation

The decay channel is followed by a buncher section that uses RF cavities in a frequency range compatible with 325 MHz in order to form the muon beam into a train of bunches, and by a phase–energy rotating section that decelerates the leading high-energy bunches and accelerates the late low-energy bunches, so that each bunch has the same mean energy. The initial RF-cavity layout assumes 0.5 m long cavities placed within 0.75 m long cells. The 2 T solenoid focusing of the decay region is continued through the buncher and the following rotator section. The RF gradient is increased from cell to cell along the buncher, and the beam is captured into a string of bunches. The gradient at the end of the buncher is 15 MV/m. This gradual increase of the bunching voltage enables a somewhat adiabatic capture of the muons into separated bunches, which minimizes phase-space dilution.

One critical feature of the muon production, collection, bunching and phase rotation scheme is the production of bunches of both signs ($\mu^+$ and $\mu^-$) at roughly equal intensities. Note that all of the focusing systems are solenoids, which focus both signs, and the RF systems have stable acceleration for opposite signs separated by a phase difference of $\pi$.

All of these issues were extensively studied within the IDS-NF, therefore designs for these systems can be adopted from that study[13].

### 2.3.3.3.4    4D Cooling Channel

The initial NuMAX does not use cooling. As illustrated in Figure 2, a "place-holder" for future cooling is provided after the phase rotator. It consists of a straight drift channel with RF cavities at zero crossing to assure longitudinal focusing. An IDS-style cooling channel will be added for NuMAX+. The cooling channel[14] consists of a sequence of identical 1.5 m long cells. Each cell contains two 0.5 m long RF cavities, with 0.25 m spacing between the cavities and 1 cm thick LiH blocks at the ends of each cavity (4 per cell). The LiH blocks constitute the energy-absorbing material for ionization cooling. Each cell contains two solenoid coils of alternating signs; this yields an approximately sinusoidal variation of the magnetic field in the channel with a peak value of $\approx 2.5$ T, providing transverse focusing with $\beta_\perp \approx 0.8$ m. The total length of the cooling section is 75 m (50 cells). Based on IDS-NF simulations, the cooling channel is expected to reduce the rms transverse normalized emittance from $\varepsilon_{N,rms}$ = 18 mm·rad to $\varepsilon_{N,rms}$ = 7 mm·rad. The resulting longitudinal emittance is $\varepsilon_{L,rms} \approx$ 70 mm/bunch. Consequently, about a factor 2 improvement of neutrino flux is expected from implementation of the 4D cooling.

### 2.3.3.3.5    Acceleration

To ensure adequate survival of the short-lived muons, acceleration must occur at high average gradient. The accelerator must also accommodate the phase-space volume occupied by the beam after the cooling channel, which is still large[15]. The need for large transverse and longitudinal acceptances drives the design of the acceleration system to low RF frequency, e.g., 325 MHz. High-gradient normal conducting RF cavities at these frequencies require very high peak-power RF sources. Hence superconducting RF (SRF) cavities are preferred. In the following we choose an SRF gradient of 15 MV/m, which has been demonstrated at 325 MHz[16], and which will allow survival of about 84% of the muons as they are accelerated to 5 GeV.

The proposed muon accelerator complex consists of a single-pass, 1.2 GeV superconducting linac with 325 MHz RF cavities that captures the large muon phase-space coming from the phase rotator (in the case of NuMAX), or the factor-of-2.5 smaller phase-space after the cooling channel (in the case of NuMAX+). The linac accelerates the muons to sufficiently relativistic energies to facilitate efficient acceleration in the RLA, while adiabatically decreasing their phase-space volume. The large acceptance of the linac requires large aperture and tight focusing. This, combined with moderate beam energies, favors solenoid rather than quadrupole focusing for the entire linac. The linac is followed by a 4.5-pass, 0.85 GeV, 325 MHz "dogbone" recirculating linear accelerator (RLA) that further compresses and shapes the longitudinal and transverse phase-space, while increasing the energy to 5 GeV. At the ends of the RLA linacs the beams need to be directed into the appropriate energy-dependent (pass-dependent) "droplet" arc for recirculation[17]. The phase space at the RLA exit is characterized by $\Delta p/p$ = 0.012 (rms) and $\Delta z$ = 8.6 cm (rms)[14].

The overall layout of the accelerator complex is shown in Figure 17.

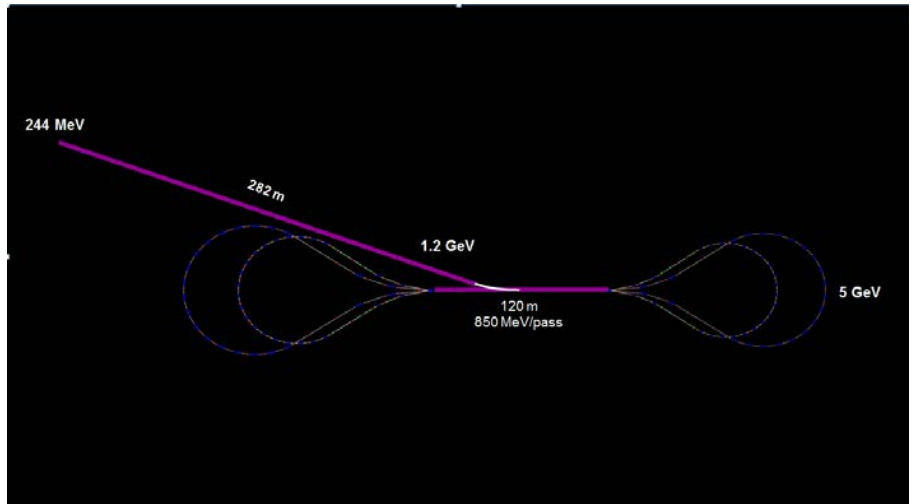

**Figure 17:** Layout of the accelerator complex: single-pass linac and 4.5-pass RLA to 5 GeV

Accelerator performance with 4D cooling as described in the previous subsection features dynamic losses of only ~ 0.5%. The same accelerator complex was recently studied for beam transport with the larger transverse emittances corresponding to the NuMAX scenario without 4D cooling. Assuming the same physical apertures in the linac and RLA, multiparticle tracking showed ~ 52% loss of the beam, thus reducing the effective muon flux by about a factor of two as compared to NuMAX+ due to the lack of 4D cooling in NuMAX.

### 2.3.3.3.6   Storage Ring

Extensive studies of muon decay rings favor a racetrack geometry, where muon of both signs can be stored in a single ring into which $\mu^+$ and $\mu^-$ bunches are injected in opposite directions, and both long straight sections point towards the same distant detector. One straight section provides a neutrino beam from $\mu^+$ decays, and the other from $\mu^-$ decays. Optimally the muon decay ring has circumference corresponding to an integral number of proton driver cycle times. One can adapt an earlier design of a 4 GeV decay ring[14], or one scaled from the IDS-NF design, with neutrino-beam-forming "production" straight sections chosen to be 235 m long and corresponding ring circumference ~ 600 m.

To minimize neutrino flux uncertainties, the rms muon beam divergence in the production straight section, $\theta_B$, must be much smaller than the rms neutrino beam divergence, $\theta_D$, arising from muon decay kinematics. The design criterion is $\theta_B < 0.1 \ \theta_D$. The fraction of muons that decay in the storage ring while traveling in the direction of the distant detector is determined by the ratio of the production straight section length to the ring circumference. With the parameters above, that ratio is about 0.4. Finally, the ring must accommodate the muon-beam momentum spread which, after acceleration, is $\Delta p/p = 0.03$. A sufficiently large momentum acceptance requires chromaticity correction through the use of sextupoles.

### 2.3.4   Required R&D

Since the initial-stage Neutrino Factory, NuMAX, relies on proton beam power of 1 MW at 3 GeV provided by the second phase of Project X with no cooling, its critical challenges are limited to:

# U.S. Muon Accelerator Program

- Proton driver and target corresponding to the state of the art in operation at SNS and therefore no specific development needed;
- A 15–20 T solenoid to efficiently capture the pions produced in the target;
- Accelerating gradient in low frequency (325–975 MHz) RF structures immersed in high magnetic field as required by the front end;
- High efficiency recirculating linear accelerators (RLA);
- 10 kt magnetized liquid argon (LAr) or magnetized fully active plastic-scintillator detector.

The high-field solenoid and RF cavities immersed in large magnetic fields are major subjects of development during the MAP Feasibility Assessment phase with results expected by or before 2018. The novel RLA technology involves multi-pass arcs based on linear combined-function magnets, which allow two consecutive passes with very different energies to be transported through the same string of magnets. Such a solution combines compactness with all of the advantages of a linear nonscaling FFAG, namely, the large dynamic aperture and momentum acceptance essential for large-emittance muon beams. The dogbone RLA with 2-pass arcs is the subject of a specific proof-of-concept electron test facility, JEMMRLA (JLab Electron Model of Muon RLA), proposed to be built and operated at Jefferson Lab. The NuMAX facility could thus be built soon after the completion of the MAP feasibility study—thus, if successful, by the end of this decade.

The full-intensity Neutrino Factory, NuMAX+, is upgraded from NuMAX by additional proton beam power on target and modest cooling of the beam emittances by a factor 2.5 in both transverse planes. Its major technical challenges therefore consist of:

- Proton driver of 3 MW at 3 GeV as provided by Project X Stage IIb and corresponding upgrade of the target possibly by adopting Hg-jet target technology whose feasibility has successfully been demonstrated by the MERIT experiment at CERN.
- Transverse cooling for which the principle is being studied in the MICE experiment at RAL with first results expected in 2015 from MICE "Step IV" (one cooling station without re-acceleration) and in 2019 from "Step VI" (full cooling cell including acceleration). As described in Sec. 2.2.6, ionization cooling at reasonable intensity ($10^8$ muons/bunch) could be further tested using the proposed nuSTORM facility as a muon source with results expected by 2022. In parallel, cooling at full Muon Collider intensity ($10^{12}$ muons/bunch) could be tested with protons in the proposed ASTA test facility at FNAL.

The NuMAX+ facility could then be progressively upgraded from NuMAX by the middle of next decade.

### 2.3.5 Technology validation for following phase

A lower-intensity Neutrino Factory, NuMAX, could be used as a long-baseline neutrino source and an R&D platform to test and validate transverse cooling (4D) at full muon intensity ($10^{12}$/pulse) as required by the full-intensity Neutrino Factory, NuMAX+. In addition, it would validate the injector complex at the 1 MW level as well as the corresponding target, front end and 5 GeV RLA.

A high-intensity Neutrino Factory can be obtained from NuMAX by upgrading the proton driver to the nominal power of 3 MW at 3 GeV as planned for Project X Stage IIb. The corresponding target and muon capture sections would need to be modified accordingly. Performance would benefit from the 4D cooling validated as R&D at NuMAX. This facility does not require any longitudinal cooling but would be used as a muon source and an R&D platform to test and validate transverse and longitudinal (6D) cooling to full specification and nominal muon bunch intensity ($10^{12}$/bunch) as required by Muon Colliders.

## 2.4   Muon Colliders

### 2.4.1   Physics specific to Muon Colliders

#### 2.4.1.1   Higgs Factory Physics

A Higgs Factory is an attractive first stage in the development of a high-energy Muon Collider. The basic design of a Muon Collider including cooling and detector concepts could be tested in full with important physics goals achieved in the process. Because the Muon Collider has the possibility of very precise energy resolution and an enhanced coupling to the Higgs ($4 \times 10^4$ larger than in an $e^+e^-$ collider), a large cross section for s-channel Higgs production is possible. For a beam spread of 4.2 MeV the Higgs cross section on resonance is 17 pb when taking into account Initial State Radiation (ISR). This allows unparalleled precision in the measurement of the Higgs mass ($\delta M = 0.10$ MeV) and a direct measurement of its width ($\delta\Gamma = 0.24$ MeV) with integrated luminosity of 100 pb$^{-1}$ [18]. Precision measurements of branching ratios are also possible. Access to second-generation Higgs couplings is assured because of the entrance channel: the branching-ratio product $\mathcal{B}(h \to \mu^+\mu^-) \times \mathcal{B}(h \to W^+W^-)$ will be determined to 2%. Precise measurement of $\mathcal{B}(h \to b\bar{b})$ is also possible and other branching ratios are under study [19]. In the (unlikely) event that the 126 GeV boson is actually a nearly degenerate doublet of states, only the MC could disentangle these states [20]. More plausible is the two-Higgs-doublet model (discussed in greater detail below) in which as the mass of the A increases it becomes more and more degenerate with the $H^0$. Here too the MC could disentangle the physics up to nearly $M_A = 900$ GeV.

In addition to direct production of a Higgs boson at a Higgs Factory, the Higgs boson can be studied in a number of other ways at a multi-TeV Muon Collider:

1) In associated production: $\mu^+\mu^- \to Z^* \to Z^0 + h^0$ has a cross section ratio R = 0.12 (see formula below). We can measure the b-quark–Higgs Yukawa coupling and look for invisible decay modes of the Higgs boson.

2) Higgsstrahlung: $\mu^+\mu^- \to t\bar{t}h^0$ has a cross section ratio R = 0.01. This process could provide a direct measurement of the top-quark–Higgs Yukawa coupling. However such a study is very challenging requiring at least 5 ab$^{-1}$ of integrated luminosity.

3) $W^*W^*$ fusion into $\bar{\nu}_\mu\nu_\mu h^0$ has a cross section ratio R = 1.1s ln(s) (for $m_h = 120$ GeV). It allows the study of Higgs self-coupling and certain rare decay modes. Its rate grows with s, an advantage for a Muon Collider.

#### 2.4.1.2   High Energy Muon Collider Physics

The Muon Collider is first and foremost an energy frontier machine. It offers both discovery, as well as precision, measurement capabilities. The physics goals of a Muon Collider are for the

---

most part the same as those of a linear electron-positron collider (ILC/CLIC)[21,22,23] at the same energy. The main advantages of a MC are the ability to study the direct (s-channel) production of scalar resonances, a much better energy resolution (because of the lack of significant beamstrahlung), and the possibility of extending operations to very high energies. At ILC/CLIC, however, significantly greater polarization of the initial beams is possible[22].

### 2.4.1.2.1    Basic processes

There are basically three kinds of channels of interest for a lepton collider: (1) open pair production, (2) s-channel resonance production, and (3) fusion processes.

- Pair Production:  The kinematic thresholds for pair production of Standard Model particles $(X^+X^-)$ are well below $E_{CM} = 500$ GeV.  For Standard Model particles at $E_{CM} > 1$ TeV the typical open-channel pair-production process is well above its kinematic threshold and the cross section becomes nearly flat in

$$R \equiv \frac{\sigma(\mu^+\mu^- \to X + \bar{X})}{\sigma_{QED}(\mu^+\mu^- \to e^+e^-)}$$

For the MC a forward/backward angular cut (e.g., $10°$) is imposed on the outgoing pair. Closer to the beam direction, a shielding cone is needed to suppress detector backgrounds arising from the effects of muon decay in the beam.

For a process whose rate is one unit of R, an integrated luminosity of 100 fb$^{-1}$ at $E_{CM}$ = 3 TeV yields 1000 events.  As an example, the rate of top quark pair production at 3 TeV is only 1.86 units of R.  This clearly demonstrates the need for high luminosity in a multi-TeV lepton collider.

- Resonances:  In addition to the $Z^0$ and Higgs resonances at low energy, many models beyond the SM predict resonances that may be produced directly in the s-channel at a high energy Muon Collider.  Here, the narrow beam energy spread of a Muon Collider, $\delta E/E \sim 10^{-3}$, could be an important advantage.  The cross section for the production of an s-channel resonance, X, with spin J, mass M and width $\Gamma$ is given by

$$\sigma_{\mu^+\mu^- \to X} = \frac{\pi}{4k^2}(2J+1)\frac{\Gamma^2}{(E-M)^2 + \Gamma^2/4}B_{\mu^+\mu^-}B_{visible}$$

where k is the momentum of the incoming muon, E the total energy of the initial system ($E_{CM}$), $B_{\mu^+\mu^-}$  $\Gamma$ the partial width of X $\to \mu^+\mu^-$, and $B_{visible}$ the visible decay width of X.  At the peak of the resonance with negligible beam energy spread,

$$R_{peak} = 3(2J+1)\frac{B(\mu^+\mu^-)B(visible)}{\alpha_{EM}^2}$$

For a sequential Standard Model $Z'$ gauge boson, the value of $R_{peak}$ is strikingly large, typically $\sim 10^4$.  The luminosity required to produce 1000 events on the $Z'$ peak with a mass

---

$s = E_{c.m.}^2$

(at the present LHC limit) of 2.5 TeV is only $3.0 \times 10^{30}$ cm$^{-2}$ sec$^{-1}$. One could envision a muon collider of significantly higher energy and reduced luminosity, suitable for the study of a Z′ with mass up to 10 TeV if a Z′ is discovered at the LHC.

Theories with extra dimensions that have a radius of curvature close to the Terascale would have Kaluza-Klein (KK) excitations. The present LHC limits on such KK modes are already well in excess of 1 TeV[24]. The lowest KK modes would be observable as resonances at a multi-TeV Muon Collider if kinematically accessible. These include the Z′ and γ′ of the electroweak sector. In theories such as the Randall-Sundrum warped extra dimensions models[25], the graviton spectrum contains additional resonances (KK-modes).

- Electroweak Boson Fusion: A typical fusion process ($\mu^+\mu^- \to \bar{\nu}_\mu\nu_\mu W^+W^- \to \bar{\nu}_\mu\nu_\mu X$) is shown in Figure 18. For $E_{CM} >> M_X$ the cross section is typically large and grows logarithmically with s = $E_{CM}^2$, while the usual pair-production processes are constant in R and thus drop like 1/s. Thus, for asymptotically high energies fusion processes dominate over standard top pair production (Figure 19). The large rates for WW, WZ and ZZ fusion processes imply that the multi-TeV Muon Collider is also effectively an electroweak-boson collider.

Physics studies of fusion processes such as $\mu^+\mu^- \to \mu^+\mu^- Z^0Z^0 \to \mu^+\mu^- X$ benefit greatly by the tagging of the outgoing $\mu^\pm$ and hence will be sensitive to the required angular cut.

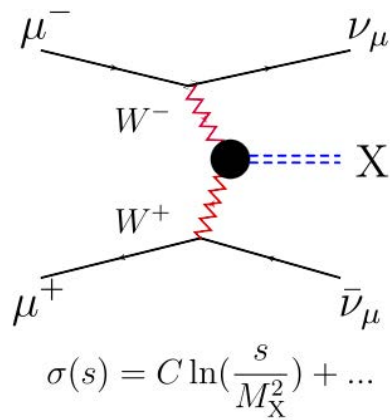

$$\sigma(s) = C\ln\left(\frac{s}{M_X^2}\right) + \dots$$

**Figure 18:** Typical boson-fusion process

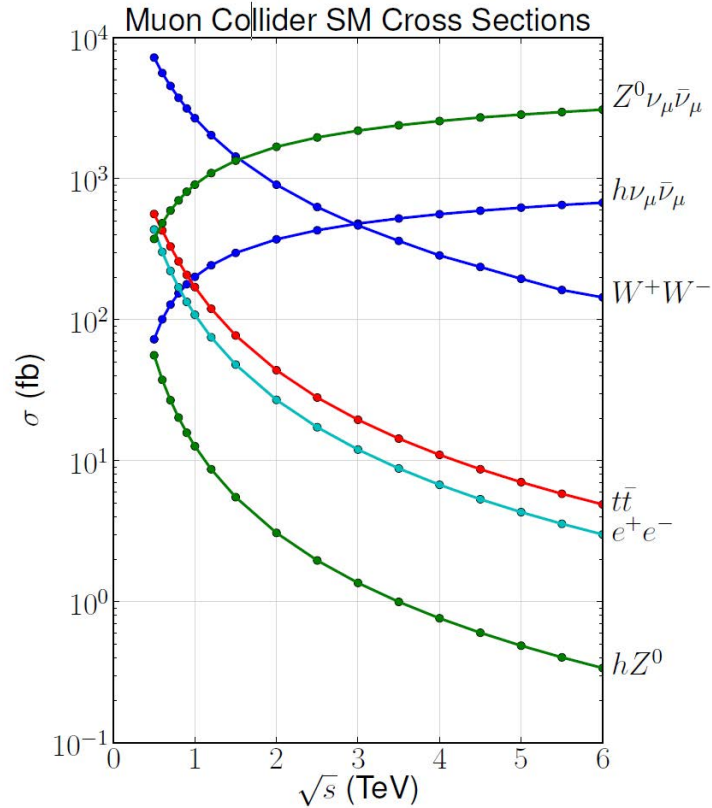

**Figure 19:** Various SM cross sections (in fb) at a MC as a function of energy from $\sqrt{s}$ = 0.5−6 TeV. The growing dominance with increasing energy of the fusion processes ($Z^0\nu_\mu\bar{\nu}_\mu$ and $h\nu_\mu\bar{\nu}_\mu$) over the standard pair production cross sections ($W^+W^-$, $t\bar{t}$, $e^+e^-$ and $hZ^0$) is clearly visible.

### 2.4.1.2.2    New Physics

#### 2.4.1.2.2.1    Extended Higgs Sector

In the two-Higgs doublet scenario there are five scalars: two charged scalars $H^\pm$, two neutral CP-even scalars h, $H^0$, and a CP-odd neutral A. For the supersymmetric MSSM models, as the mass of the A is increased, the h becomes closer to the SM Higgs couplings and the other four Higgs become nearly degenerate in mass ("decoupling"). This makes resolving the two neutral-CP states difficult without the good energy resolution of a Muon Collider. This separation in the case of $M_A$ = 400 and tan β = 5 was studied in detail by Dittmaier and Kaiser[20]. The Muon Collider is an ideal place to study s-channel production of very heavy H/A because of decoupling[26]. This is a typical situation in SUSY models that evade the LHC bounds. A comparison of associated-production mechanisms for heavy-Higgs production (available at both ILC/CLIC and an MC) with the s-channel production available only at an MC is shown in Figure 20. The resonance production cross section is more than two orders of magnitude larger than that of any process available at CLIC.

---

[26] E. Eichten and A. Martin, "The Muon Collider as a H/A Factory," arXiv:1306.2609 [hep-ph]

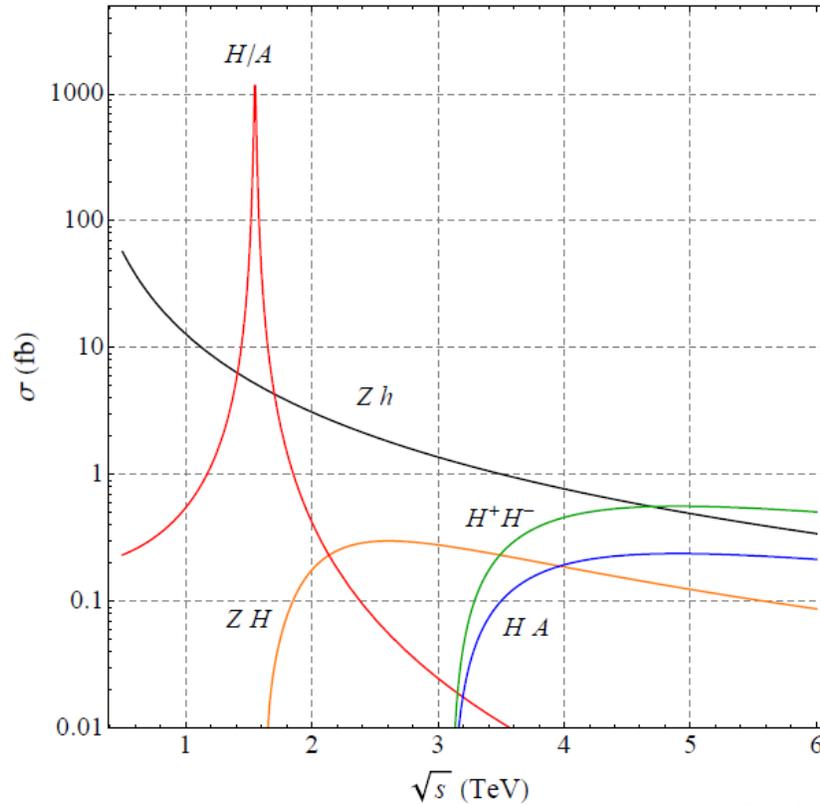

**Figure 20:** Comparison of H/A resonance production at a Muon Collider with $Z^0$h, $Z^0$H and heavy Higgs pair production common to both a Muon Collider and an $e^+e^-$ Linear Collider. The SUSY model is an ILC benchmark Natural Supersymmetry model with $m_A$ = 1.55 TeV, $\Gamma_A$ = 19.2 GeV and $m_H$ = 1.56 TeV, $\Gamma_H$ = 19.5 GeV. In spite of the near degeneracy of the H/A resonances (combined here), properties of each individual state and its decay modes can be disentangled at a Muon Collider.

### 2.4.1.2.2.2    Supersymmetry

Supersymmetry (SUSY) provides a solution to the naturalness problem of the SM. It is a symmetry that connects scalars with fermions, ordinary particles with superpartners—a symmetry that is missing in the SM.

The simplest SUSY model is the cMSSM, with only five parameters determining the masses of all the superpartners. It is now highly constrained by direct limits on the Higgs, mainly from LEP, CDF and DZero. Z-pole studies have provided constraints from electroweak precision measurements, and we have no indication of SUSY from flavor physics so far[27]. Recently the LHC has produced strong lower bounds on the masses of squarks and gluinos[28,29]. All this, taken together, makes it almost certain that direct coverage of the remaining MSSM parameter space requires a multi-TeV scale lepton collider such as CLIC (at 3 TeV) or a Muon Collider. Because the production cross sections of pairs of supersymmetric particles are generally smaller than the pair production for SM particles (see Figure 21) high luminosity is required at such a multi-TeV collider (1 ab⁻¹).

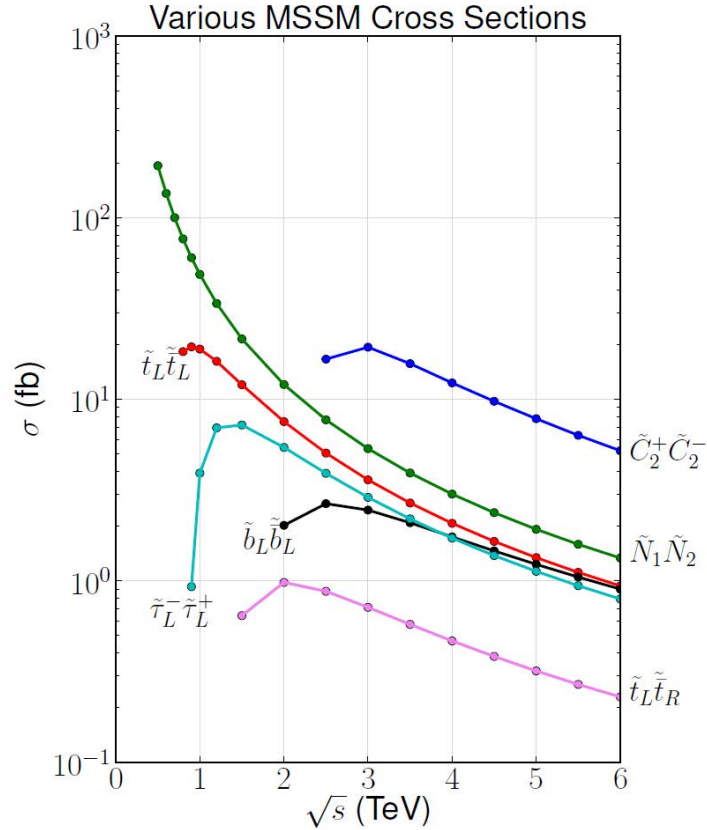

**Figure 21:** Various SUSY model cross sections (in fb) at a multi-TeV MC vs energy ($\sqrt{s} = 0.5 - 6.0$ TeV) in the CSS 2013 benchmark Natural SUSY model.

### 2.4.1.2.2.3    New Strong Dynamics

The discovery of a Higgs-like boson at 126 GeV was unexpected in strong dynamical models of electroweak symmetry breaking. However strong dynamics may still play a role in the solution of the naturalness problem of the Standard Model[30]. For details and a discussion of various new strong dynamics models (e.g., Topcolor, TC2, and Light Higgs models) see the review of Hill and Simmons[31]. There have even been suggestions that the Higgs boson is a dilaton, i.e., a composite state associated with a nearly conformal strong dynamics at a nearby scale[32]. All these models generally predict new strongly interacting particles in the few-TeV range and would provide a rich spectrum of states that can be observed at a multi-TeV Muon Collider.

# U.S. Muon Accelerator Program

### 2.4.1.2.2.4    Contact Interactions

New physics can enter through contact interactions, which are higher-dimension four-fermion operators in the effective Lagrangian ($(4\pi/\Lambda^2)(\bar{\Psi}\Gamma\Psi)(\bar{\Psi}\Gamma\Psi)$). The MC is sensitive to $\Lambda \sim 200$ TeV, roughly equivalent to CLIC. Preliminary studies suggest that the forward angle block-out is not an issue here[28]. If polarization is not available at an MC, the MC may be at a disadvantage compared with CLIC in being able to disentangle the chiral structures of the new operators.

### 2.4.1.2.3    Summary

A multi-TeV Muon Collider is required for the full coverage of Terascale physics. The physics potential for a Muon Collider at $\approx 3$ TeV and integrated luminosity of 1 ab$^{-1}$ is outstanding. Particularly strong cases can be made if the new physics is SUSY or new strong dynamics. Furthermore, a staged Muon Collider can provide a Higgs Factory with unique abilities as well as a Neutrino Factory to fully disentangle neutrino physics. If narrow s-channel resonance states exist in the multi-TeV region, the physics program at a Muon Collider could begin with less than $10^{31}$ cm$^{-2}$ sec$^{-1}$ luminosity.

Detailed studies of the physics case for a 1.5 TeV Muon Collider are in the early stages. The goals of such studies are to: (1) identify benchmark physics processes; (2) study the physics dependence on beam parameters; (3) estimate detector backgrounds; and (4) compare the physics potential of a Muon Collider with those of the ILC, CLIC and upgrades to the LHC.

## 2.4.2    The detector in a phased approach

### 2.4.2.1    Physics Environment

Muons have distinct advantages as projectiles in a colliding-beam accelerator. They are point-like, so one can adjust the center-of-mass energy of the collision precisely and study resonance structures and threshold effects in great detail. Muons are 207 times more massive than electrons, therefore they radiate ~$10^4$ times less than electrons traveling with the same radius of curvature and momentum. A high-energy Muon Collider can therefore be built as a circular, rather than a linear, machine. In addition, beamstrahlung effects—radiation due to beam-beam interactions—are much smaller at a Muon Collider than at an $e^+e^-$ machine, allowing precise beam-energy constraints. Two or more interaction regions at a Muon Collider provide an opportunity to perform multiple experiments simultaneously. Finally the mass-dependent coupling of the Higgs boson to the $\mu^+\mu^-$ system is 40,000 times larger than the coupling to $e^+e^-$, making a Muon Collider an ideal candidate for direct study of Higgs bosons produced in the s-channel and lowering the center-of-mass energy of the Higgs Factory Muon Collider compared to that of an $e^+e^-$ machine.

### 2.4.2.2    s-Channel Higgs Factory

In the s-channel Higgs Factory, a Muon Collider with a $4 \times 10^{-5}$ beam energy spread provides an opportunity to generate up to 13,500 Higgs bosons per year ($10^7$ s) and to study their properties with unprecedented precision. The major differences in physics environment between a Muon Collider and an $e^+e^-$ machine that impact detector design are:

# U.S. Muon Accelerator Program

- Lower beamstrahlung in a Muon Collider, enabling more effective beam constraints and sharper distributions for physics signals;
- Typically smaller levels of beam polarization: 15% muon vs 80% electron polarization;
- Beam shielding required in a Muon Collider limits acceptance in the forward direction.

The radiation environment in a Muon Collider is similar to that at LHC, which will require detectors with moderate radiation hardness.

Muon Collider beam energy can be measured with a precision better then $10^{-5}$ by utilizing the g-2 spin precession of beam muons[33]. With beam energy spread similar to the predicted 4.2 MeV width of the Higgs a model-independent measurement of the Higgs width could be the unique, flagship measurement of such a machine. With straightforward event shape cuts the Higgs → $b\bar{b}$ signal/$\sqrt{\text{background}}$ ratio can be close to 3[34]. A beam energy scan with 1 fb$^{-1}$ integrated luminosity, counting the Higgs yield as a function of the center-of-mass energy, can establish the mass of the Higgs to a statistical precision better than 0.1 MeV and the width to better than 0.5 MeV[35] as shown in Figure 22. Here the crucial factors are establishment, measurement, and maintenance of a small beam energy spread and precise monitoring of the beam energy. Figure 23 shows the cross section of a possible Higgs Factory Muon Collider detector consisting of precise tracking, calorimetry and muon detection. Shielding of detectors from beam-induced radiation is discussed later in this section.

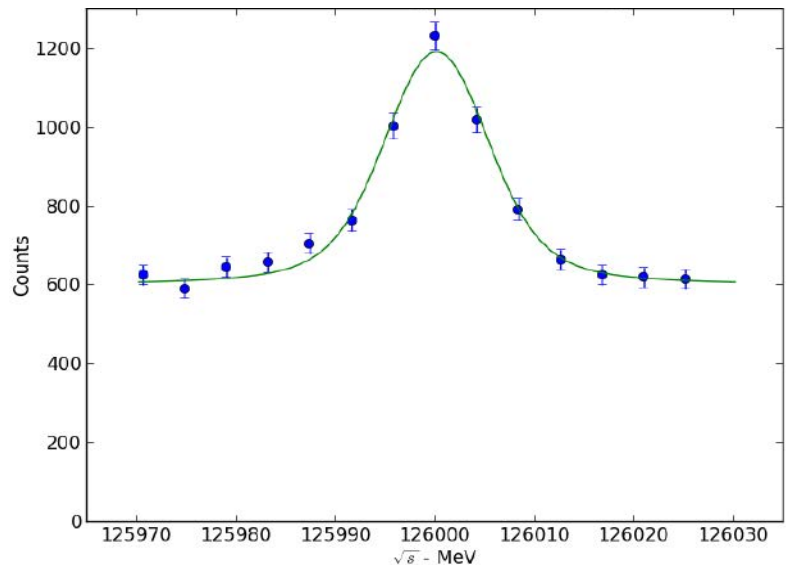

**Figure 22:** Simulated $b\bar{b}$ event counts from a 1 fb$^{-1}$ scan across a 126 GeV Higgs peak assuming 4.2 MeV beam energy spread.

# U.S. Muon Accelerator Program

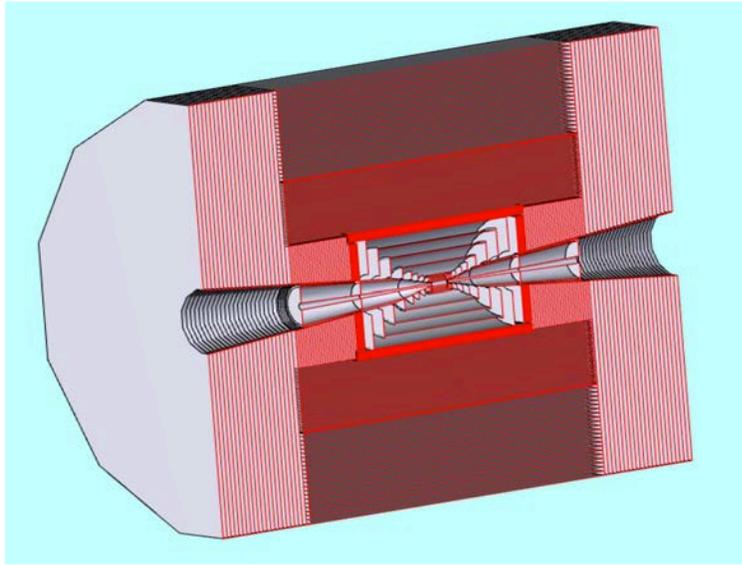

**Figure 23:** Cross sectional view of a possible Higgs Factory Muon Collider detector showing the tungsten cones shielding the detector from beam related backgrounds.

### 2.4.2.3    *High Energy Muon Collider*

The physics goals of a high-energy (~3 TeV) Muon Collider would be similar to those of a high-energy $e^+e^-$ collider such as ILC or CLIC. These machines would be intended to search for and make precision measurements of new physics. The detector requirements for a number of beyond-Standard Model scenarios such as $Z'$, supersymmetry, extra dimensions, and composite Higgs have been studied in the ILC DBD documents[36] and CLIC TDR[37]. Performance goals for track momentum resolution, impact parameter resolution, and jet energy resolution for a Muon Collider are shared with those of ILC or CLIC. There are also some differences. For example, if we assume that the 126 GeV Higgs mass and width will be measured by an s-channel Higgs Factory, the constraints on tracker momentum resolution, which in the ILC are driven by Higgs recoil mass measurements, are not as severe at a Muon Collider as at the ILC.

In the tracker, the inner layer of the vertex detector might need to be at a larger radius than at the ILC in order to avoid the halo accompanying the muon beam. Additional mass might be needed for tracker cooling due to the power required for fast timing electronics and the low operating temperatures needed for radiation hardness. Calorimetry for a high-energy Muon Collider will become even more precise due to the high energy depositions and will be limited by the stability of calibration and the constant term. The muon detector for a Muon Collider (Figure 24) is likely to be similar to those of modern collider detectors, such as at the Tevatron and LHC.

---

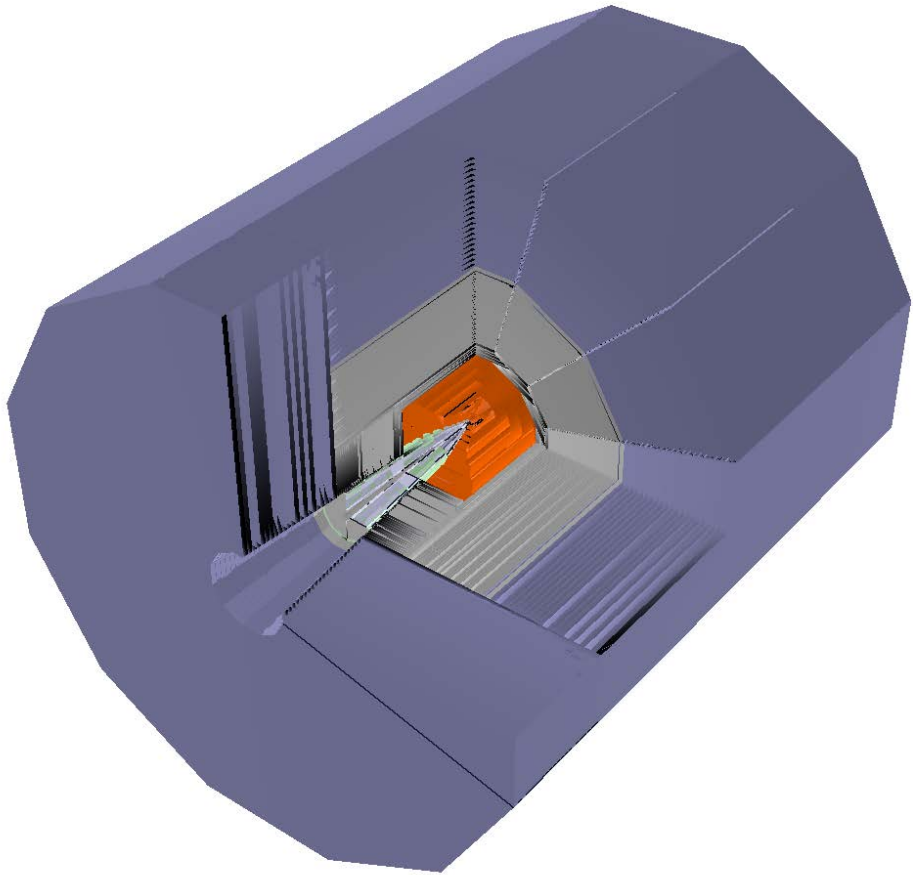

**Figure 24:** Illustration of the Muon Collider detector

### *2.4.2.4   Backgrounds*

#### 2.4.2.4.1   Background Sources

The primary source of detector backgrounds in a Muon Collider detector is the $\mu \rightarrow e\nu$ decay of the muon beam. The decay electrons are transported around the collider ring until they interact with beamline components or in one of the masks designed to absorb beam backgrounds. The subsequent electromagnetic showers generate a sea of low-energy photons and soft neutrons that enter the detector region at all radii. A significant fraction of the background is generated by beam interactions many meters upstream of the interaction point.

There is also a halo of decay electrons accompanying the beam to the interaction point. If there were no specifically designed shielding this halo would produce an unacceptable level of electromagnetic background in the detector. For this reason a tungsten cone lined with borated polyethylene surrounds the interaction point. This cone reduces the electromagnetic energy entering the detector volume by two orders of magnitude. The cone angle is about 10° for a high-energy collider and increases to 15° for a Higgs Factory. The larger angle in the Higgs Factory case allows for the large aperture and strong focusing needed to achieve high luminosity.

# U.S. Muon Accelerator Program

### 2.4.2.4.2    Background Simulation

The intimate coupling of the background and beam transport means that backgrounds must be simulated as part of the accelerator beam transport as well as in the local machine–detector interface. This background generation has been implemented in the MARS and G4beamline program frameworks. Background simulation is complete for the 1.5 TeV Muon Collider design, which has a complete lattice design available. The Muon Collider Higgs Factory lattice has recently been designed and full background simulations are becoming available.

Figure 25 shows arrival times of energy deposition in the detector volume for a 1.5 TeV collider. Most of the background consists of low-energy photons and neutrons that are significantly delayed with respect to the particles emerging from the interaction point.

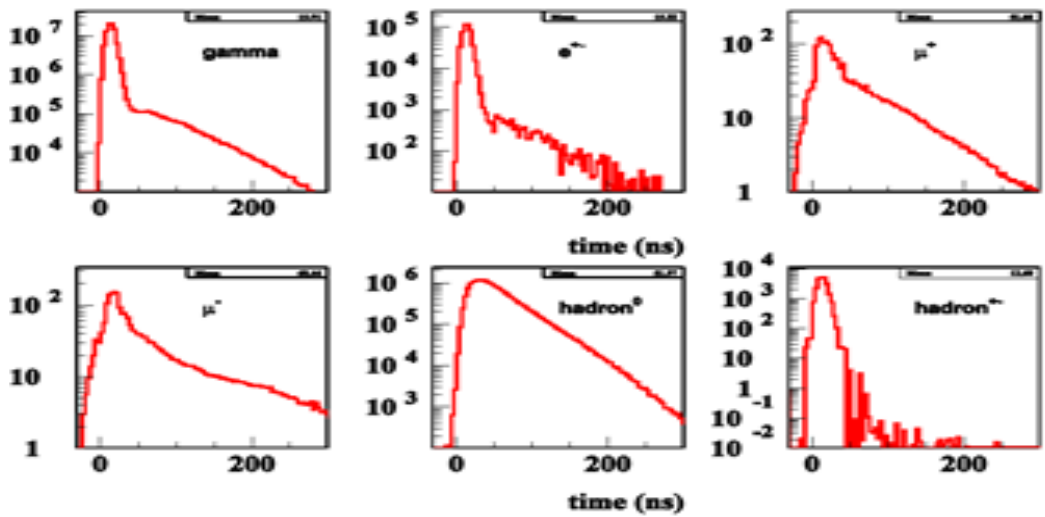

**Figure 25:** Arrival time for various beam background components in a 1.5 TeV Muon Collider

### 2.4.2.4.3    Background reduction

The fact that much of the background is low-energy and out-of-time provides key handles for background reduction. Figure 26 shows a plot of track path length within a silicon detector in a barrel silicon layer as a function of dE/dx. Soft photons from the background can Compton scatter, producing electrons with short range and moderate energy loss, or convert. Neutrons typically scatter with small energy loss. A moderate dE/dx cut can discriminate against the bulk of the neutron background. The most powerful discriminator against background is timing. Typically fewer than 1 in 1000 background hits pass a timing cut requiring the hits to be consistent with a prompt signal from the vertex[38].

---

[38] N. Terentiev *et al.*, Phys. Procedia **37**, 104 (2012).

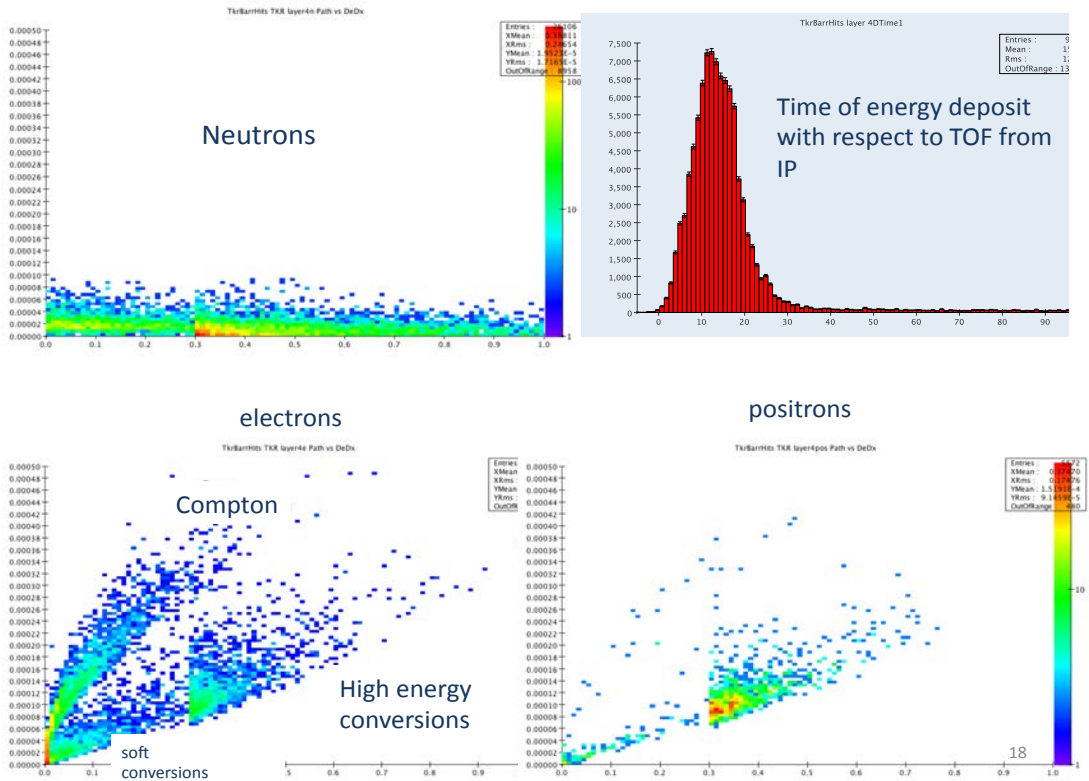

**Figure 26:** Contributions of various background components to signals in a barrel silicon detector layer

Calorimetry will also utilize timing to reduce beam backgrounds. This is effective for the prompt electromagnetic showers from primary electrons and photons. Hadronic showers are more complex, with a significant amount of energy arriving late due to hadronic interactions and nuclear de-excitation[39]. A timing cut may therefore degrade the hadron calorimeter energy resolution. Jet energy resolution of 5% is needed for 2σ discrimination between W and Z decaying to a pair of jets. Calorimetry schemes based on total absorption dual readout technology as well as a highly pixelated digital calorimeter are being studied and show promise in providing the required resolution including timing cuts[40].

### 2.4.3   The facility description in a phased approach

Here we focus on the staging of modifications and additions to each of the facilities described above. The staged Neutrino Factory (NuMAX → NuMAX+) would pave the path for a future Higgs Factory and high-energy Muon Collider. After the Neutrino Factory, the proposed scenario envisions deployment of a Higgs Factory, which could potentially begin commissioning and scanning operation using the proton beam provided by Stage II of Project X and then upgrade to operation with a 4MW beam. This would be followed by the deployment of a "Final Cooling" channel, additional acceleration stages and a larger collider ring for a TeV-scale Muon Collider

---

[39] F. Simon, arXiv:1109.3143.
[40] R. Raja, JINST **7** (2012) P0401.

# U.S. Muon Accelerator Program

(MC), thus providing the final elements of a Muon Accelerator Staging Plan which spans the Intensity and Energy Frontiers—in a nutshell,

- nuSTORM → NuMAX → NuMAX+ → HF(commissioning) → HF(operation) → TeV-scale MC

## 2.4.3.1 Components

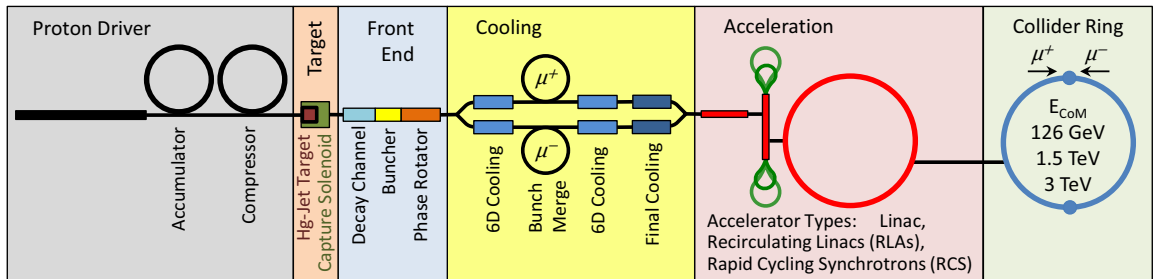

**Figure 27:** Functional elements of a Higgs Factory/Muon Collider complex

The functional elements of a Higgs Factory/TeV-scale Muon Collider complex are illustrated schematically in Figure 27. They can be listed as follows:

- A proton driver producing a high-power multi-GeV bunched proton beam.

- A pion production target operating in a high-field solenoid. The solenoid confines the pions radially, guiding them into a decay channel.

- A "front end" consisting of a solenoid π→μ decay channel, followed by a system of RF cavities to capture the muons longitudinally and phase rotate them into a bunch train suitable for use in the cooling channel.

- A cooling channel that uses ionization cooling to reduce the longitudinal phase space occupied by the beam by about six orders of magnitude from the initial volume at the exit of the front end. The first stages of the cooling scheme include 6D cooling and a bunch merge section. For a Higgs Factory, cooling would stop before entering a "Final Cooling" section which trades increased longitudinal emittance for a ten-fold improvement in each transverse emittance as required for a high luminosity TeV-scale Muon Collider.

- A series of acceleration stages to take the muon beams to the relevant collider energies. Depending on the final energy required, this chain may include an initial linac followed by recirculating linear accelerators (RLA) and/or fixed-field alternating gradient (FFAG) rings. At present, the multi-TeV collider designs utilize rapid-cycling synchrotrons (RCS) as the baseline for achieving the highest beam energies.

- A compact collider ring, having a circumference of ~300 m for a Higgs Factory and several kilometers for a TeV-scale collider, along with the associated detector(s). At present, the baseline Higgs Factory design assumes 1 detector while the TeV-scale colliders can readily accommodate at least 2 detectors.

## 2.4.3.2 Implementation on the Fermilab site

Here we discuss specific facilities based on Fermilab's infrastructure and integrated with the stages of Project X. Based on the physics needs identified at the time, the facility could support

# U.S. Muon Accelerator Program

in succession a Higgs Factory and/or TeV-scale collider in the 1–10 TeV range. The parameters for the MAP baseline configurations are summarized in Table 2. A layout of the facility on the FNAL site (including nuSTORM, NuMAX and a Higgs Factory) is given in Figure 2.

The collider implementation builds directly on the foundation of the previously described Neutrino Factory subsystems. In particular, the proton driver, target, front end and acceleration systems will all be reused, with incremental modifications in some cases. Specific changes for the evolution from NuMAX to the HF would include:

- An additional compressor after the accumulator ring to obtain the necessary bunch parameters for collider operation.
- Upgrade of the 4D cooling channel required for NuMAX+ operation to a full 6D cooling channel. This modification would not only provide the basis for HF operation but could provide further improvements in NuMAX+ performance.
- In order to achieve the 63 GeV/beam required for the HF, the initial acceleration chain could be reused. It could be followed by a 9-pass RLA based on a 650 MHz (or possibly 1300 MHz) SCRF system. The 9-pass RLA would utilize the recently developed 2-pass arc design[41] as shown in Figure 2. Alternatively, an RCS-based design has also been explored.
- For a Higgs Factory, a 300 m circumference collider with a single detector is envisioned.

In principle, initial commissioning of the HF could begin with the beams provided by Stage II of Project X. Nominal operation assumes that the 4 MW upgrades to Project X and the target system are in place. Table 2 summarizes the expected performance of the configuration available during a short (estimated as 1.5–2 year) commissioning phase along with the parameters of the fully operational machine with a baseline luminosity of $0.8 \times 10^{32}$ cm$^{-2}$ s$^{-1}$. Such a machine offers a rate of Higgs production that is quite competitive with that of the proposed linear collider. Furthermore it offers the exquisite energy resolution needed in order to measure the width of a single Higgs directly, or explore a more complicated dual-Higgs scenario if nature proves more complex. Furthermore, more advanced muon cooling concepts are presently being explored which offer the potential to increase the HF performance further by reducing both the longitudinal and transverse emittances of the beams supplied to the collider. While these concepts are not yet sufficiently validated to be included in the HF baseline, they could increase the luminosity to several $\times 10^{32}$ cm$^{-2}$ s$^{-1}$ and also provide further improvements to the energy resolution.

An evolution of the facility to support a multi-TeV Muon Collider is straightforward if required in order to complete the physics landscape. This would require addition of:

- A final 6D cooling section to further reduce both transverse emittances by a factor 10.
- Additional acceleration by rapid-cycling synchrotrons to rapidly and efficiently reach the desired beam energy.
- Addition of a suitable collider ring which could accommodate (at least) 2 detectors. The baseline designs for rings at center-of-mass energies of 1.5 and 3 TeV are provided in Table 2, and higher-energy rings have also been considered. The 1.5 (3) TeV design has a circumference of 2.5 (4.5) km.

---

[41] V.S. Morozov, S.A. Bogacz, Y.R. Roblin, K.B. Beard "Linear Fixed-field Multipass Arcs for Recirculating Linear Accelerators", Phys. Rev. ST Accel. Beams **15**, 060101 (2012).

# U.S. Muon Accelerator Program

### 2.4.4    Required R&D

A Higgs Factory or a Muon Collider relies on the proton driver, front end and pre-acceleration developed for a Neutrino Factory with additional challenges, especially:

- For a Higgs Factory:
    - High-power proton linac and target station (4 MW) as foreseen in phase IV of project X although full power capability is not required for initial Higgs Factory operation;
    - Ionization cooling with reduction of 6D emittance by 6 orders of magnitude (2 in transverse plane, 4 in longitudinal plane);
    - Rapid-cycling synchrotron or FFAG rings for fast acceleration;
    - Collider ring and machine–detector interface including absorbers of radiation emitted by muon decays;
    - Detector in a high-background environment.

As developed in sections 2.2.6 and 2.3.4, ionization cooling after demonstration of its principle in the MICE experiment at RAL could be further tested and validated at reasonable intensity ($10^{10}$ muons/pulse) by using the proposed nuSTORM facility as a muon source integrating a specific R&D platform with results expected by 2022.  In parallel, cooling at full intensity ($10^{12}$ muons/bunch) could be tested with protons in the proposed ASTA test facility at FNAL

The FFAG concepts are the subject of a specific test facility, EMMA, in operation at Daresbury Laboratory in the UK.

A Higgs Factory could therefore be envisaged by the middle of next decade.

- For a TeV-scale Muon Collider (assuming a Higgs Factory or low-energy collider is previously built):
    - Very high field (> 30 T) solenoids utilizing high-temperature superconducting coils as required in the final cooling section in order to damp the transverse emittances for high-luminosity operation;
    - Rapid-cycling synchrotron or FFAG rings for further fast acceleration;
    - Collider ring and machine–detector interface including absorbers of larger amount of radiation emitted by muon decays;
    - Detector operation in a higher-background environment.

The last three items will benefit from operational experience at the low-energy collider such that the only major additional development concerns the very high field solenoids, already one of the major subjects of the present MAP technology feasibility study.

### 2.4.5    Technology validation for following phase

A low-energy Muon Collider (Higgs Factory) could be upgraded in luminosity by increasing the proton power on target and/or improving the transverse cooling while preserving the longitudinal emittance.  It represents a logical step towards a multi-TeV collider which would reuse a number of its systems, especially:

- The proton driver injector complex including the proton linac, accumulator and compressor.
- The high-power target.
- The front end including the decay channel, buncher and phase rotator.
- The 6D ionization cooling stages.
- The pre-acceleration up to 4 GeV.

It would be upgraded in energy by further RLA or RCS. It could be used as a muon source and an R&D platform to test and validate the final cooling to small transverse emittances at nominal muon bunch intensity ($10^{12}$/bunch) as required by a high-energy Muon Collider. Such a facility could be launched by 2025.

A high-energy Muon Collider will use all technologies tested and validated during the previous phases and extending further in energy by installing more RLA and/or RCS and a corresponding collider ring.

## 3    Overall schedule and Conclusion

A staging approach for muon-based facilities has been developed with physics interest and technology validation at each phase taking advantage of the present and proposed facilities at Fermilab. Each stage is built as an addition to the previous stages, reusing as much as possible of the systems already installed such that the additional budget of each stage remains affordable. Thanks to the great synergies between Neutrino Factory and Muon Collider technologies, these facilities are complementary and allow capabilities and world-leading experimental support spanning physics at both the Intensity and Energy Frontiers. Rather than building an expensive test facility without physics use, the approach uses each stage as an R&D platform at which to test and validate the technology required by the following stage. The critical issues of this novel and promising technology are thus all addressed in the most efficient and practical way within reasonable funding and schedule constraints.

As developed further in the Appendix and illustrated in Figure 1, such a staging approach with integrated R&D would allow informed decisions by 2020 about Neutrino Factories at the Intensity Frontier and by 2025 about Muon Colliders at the Energy Frontier.

# U.S. Muon Accelerator Program

## Appendix A: A response to the questions posed by the Lepton Colliders sub-group of the CSS2013 Frontier Facilities Group

### A-1: Higgs Factory

*What are the required parameters and key characteristics of lepton / gamma colliders in the Higgs factory range? With physics capabilities far beyond the LHC? What are the possible configurations of a lepton collider Higgs factory that would use existing accelerator infrastructures? How might such a facility be upgraded in energy and / or luminosity? How does a Higgs factory scale cost-wise to a TeV scale linear collider?*

The key beam parameters of a muon-based facility, ranging from a 126 GeV Higgs Factory to a multi-TeV collider, are summarized in Table 2 of the Executive summary. The Higgs Factory takes advantage of the s-channel resonance specific to muons with a cross section 40,000 times larger than for electron-positron collisions. As a consequence, the required luminosity to produce 13,500 Standard Model Higgs events during a typical $1 \times 10^7$ sec operating year is only $8 \times 10^{31}$ cm$^{-2}$s$^{-1}$. This can be compared with luminosities in the $10^{34}$ cm$^{-2}$s$^{-1}$ range, which are required to provide similar numbers of Higgs events with an electron-positron collider via associated production. In order to probe the narrow s-channel resonance, the rms beam momentum spread should not be larger than a few $\times 10^{-5}$, which requires a small longitudinal emittance and a collider ring with excellent beam energy stability and corresponding control of the injection energy. A plot of the emittance reduction through the planned muon ionization cooling channel is shown in Figure A-1. In order to achieve the small longitudinal momentum spread required for a Higgs Factory, the cooling process will stop at the end of the 6D cooling system. The final cooling section, which trades off increased longitudinal emittance to obtain the smaller transverse emittances required for a TeV-scale MC, will not be employed.

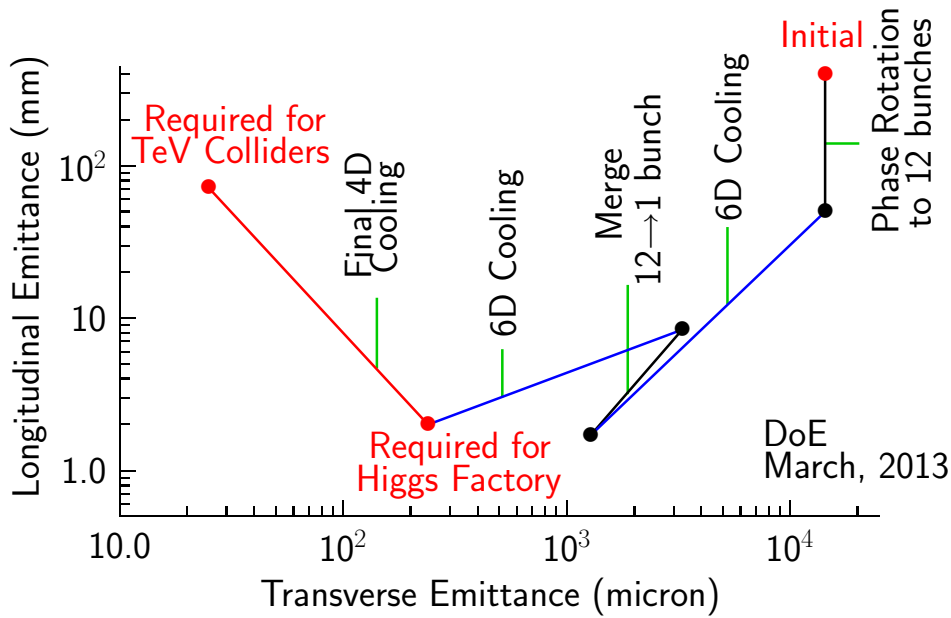

**Figure A-1:** Evolution of transverse and longitudinal beam emittance during ionization cooling

# U.S. Muon Accelerator Program

Due to the narrow momentum spread, such a facility is ideally suited to measure the Higgs boson mass and width with sub-MeV precision. A Higgs Factory, if based at FNAL, would take advantage of then-existing facilities, in particular:

- Project X Stage II (with 1–3 MW beams at 3 GeV) initially and subsequently Stage IV (with 4 MW beams at 8 GeV) as the proton driver, including a target to produce the muons as tertiary particles.
- nuSTORM, and possibly a subsequent Neutrino Factory, which would serve as a muon source for an integrated full-scale test facility. Such a facility would enable critical systems demonstrations required for a Higgs Factory, in particular a 6-dimensional muon cooling channel with high intensity beams, to be fully validated.

The Higgs Factory could be upgraded in luminosity by increasing the proton power on target and/or improving the transverse cooling while preserving the longitudinal emittance. It represents a logical stage towards a multi-TeV collider which would reuse a number of its systems, in particular:

- The proton driver injector complex including the proton linac, accumulator and compressor;
- The high-power target;
- The front end, including the decay channel, buncher and phase rotator;
- The 6D ionization cooling stages;
- The initial stages of the acceleration chain.

## A-2: TeV-Scale Collider

*What are parameters of ~TeV scale lepton colliders? Which technical approaches are naturally linked and how can they be used in conjunction with each other? How does performance scale with energy? What are the most important luminosity limitations?*

The key beam parameters of a muon-based multi-TeV collider are summarized in the Table 2. A Muon Collider is the ideal technology for a collider in the TeV or multi-TeV range. Indeed, it suffers neither from synchrotron radiation losses, as do circular colliders, nor from beamstrahlung, which results in a significant deterioration of the luminosity spectrum for linear colliders. In addition, its ability to support multiple interaction regions multiplies the total delivered luminosity, thus serving a large High Energy Physics community. Due to the beam circulating about 1000 turns before decay in the collider ring, the necessary beam emittances and beam dimensions at collision are greatly relaxed with respect to linear collider parameters.

The increase in luminosity that is obtained for a multi-TeV collider very closely approaches the ideal $E^2$ scaling, as shown in Figure A-2. The primary performance limitations at high energy will be the limits imposed by the production, cooling and lifetime of the muon beam. The ultimate colliding-beam energy is limited by the neutrino radiation at ground level and, for present design assumptions, limits the center-of-mass collider energy to $\leq 10$ TeV.

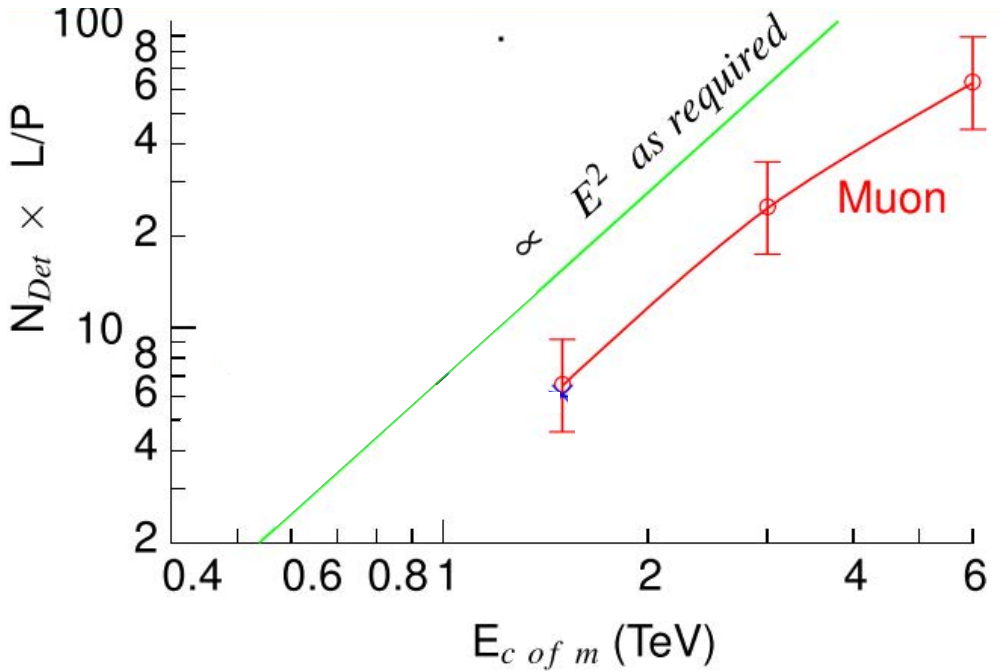

**Figure A-2**: Effective scaling of the luminosity per input power with energy for a multi-TeV MC

## A-3: Timescale and Limitations

*Can muon colliders have a role as a Higgs factory on a 10 to 15 year horizon? What are the limitations due to site-boundary radiation control and collider background for a TeV-scale muon collider?*

Muon Colliders rely on novel technologies whose feasibility is being studied in two phases culminating late this decade with the conclusion of the MAP Feasibility Assessment. If successful, a conceptual and subsequent technical design could then be launched along with a program of advanced systems R&D.

The most significant technical feasibility issue on the critical path is the design and performance of an ionization cooling channel. The basic principles of 4D (needed for the NuMAX+) and 6D (needed for a collider) ionization cooling are being studied in the Muon Ionization Cooling Experiment (MICE) at Rutherford Appleton Laboratory (RAL) in the UK with Step IV (one cooling station without acceleration) to be completed by 2016 and Step VI (one full cooling cell including acceleration) by the end of the decade. We anticipate that MICE results, in combination with the MAP Feasibility Assessment, will enable an informed decision on Neutrino Factory capabilities by 2020.

Advanced R&D for the high-intensity 6D ionization cooling channel required for a Muon Collider could be pursued using a facility such as nuSTORM to provide a muon source with significant intensity (~10$^{10}$ μ/pulse in a 100–300 MeV/c momentum slice) and in the FNAL/ASTA facility with protons at intensities of 10$^{12}$–10$^{13}$/bunch for the study of collective effects with possible results on the timescale of the early 2020s. These results, in combination with the output from the conceptual/technical design report effort, would enable an informed decision about a Muon Collider by around 2025.

An initial Neutrino Factory at Fermilab meeting the NuMAX specification would provide an enhanced physics path for a detector based at SURF. It would also provide a platform for full-scale systems validations in preparation for executing either a full power NuMAX+ or collider option. This would include implementation of 6D muon cooling at nominal beam intensities.

An initial collider operating at the Higgs resonance could be commissioned and provide useful physics (producing several thousand Higgs bosons per year) using the proton power provided by Project X Stage II (1–3 MW at 3 GeV). The eventual upgrade of the proton driver to 4 MW at 8 GeV would then enable the Higgs collider to produce upwards of 13,500 Higgs bosons per year. The next steps in the upgrade path would be to deploy the final cooling stage, higher energy accelerators and ring required for multi-TeV collider operation.

The ultimate energy of a Muon Collider is limited by neutrino radiation at ground level by muon decay. The level of radiation is maintained below 1/10 of the federal public limit by building the collider deep underground and beam scanning with a dipole so as to spread the emission cone. With such measures, the ultimate center-of-mass colliding-beam energy would be limited to about 10 TeV with a depth of 500 m (or 6 TeV without beam scanning).

## A-4: Power

### For colliders from the Higgs to TeV scale, what is the characteristic power consumption and how does it scale with energy and luminosity?

Even though muon decay requires regular replacement of the beams after roughly 1000 turns during collider operation, the average power of the beams after acceleration is typically 1.2 MW for a Higgs Factory and 11.5 MW for a 3 TeV collider. The wall-plug power for beam acceleration by a superconducting recirculating linac (RLA) at low energy and one or more rapid-cycling synchrotrons (RCS) at higher energy is therefore very reasonable. Nevertheless, significant power is required for muon beam generation and cooling. A typical partition of the power requirements is shown in Table A-1.

If 70 MW of power for cooling, services and control is included, as has been estimated in the linear collider power studies, the wall-plug power ranges from approximately 200 MW for a Higgs Factory to 230 MW for a 3 TeV collider. This compares quite favorably with the linear collider options in the same energy as shown in Figure A-3. A Muon Collider is therefore quite efficient at high energy with an attractive wall-plug power consumption increase per energy of 10 MW/TeV. Moreover, a useful figure of merit, defined as the total luminosity divided by the wall-plug consumption of the whole facility, demonstrates that the Muon Collider is an ideal technology at high colliding-beam energy because of its limited power consumption and the ability to support at least two interaction points as shown in Figure A-4.

**Table A-1**:  Wall-plug power distribution in a 1.5 TeV Muon Collider

| | Len | Static | Dynamic | — | — | — | Tot |
|---|---|---|---|---|---|---|---|
| | | 4° | rf | PS | 4° | 20° | |
| | m | MW | MW | MW | MW | MW | MW |
| p Driver (SC linac) | | | | | | | (20) |
| Target and taper | 16 | | | 15.0 | 0.4 | | 15.4 |
| Decay and phase rot | 95 | 0.1 | 0.8 | | 4.5 | | 5.4 |
| Charge separation | 14 | | | | | | |
| 6D cooling before merge | 222 | 0.6 | 7.2 | | 6.8 | 6.1 | 20.7 |
| Merge | 115 | 0.2 | 1.4 | | | | 1.6 |
| 6D cooling after merge | 428 | 0.7 | 2.8 | | | 2.6 | 6.1 |
| Final 4D cooling | 78 | 0.1 | 1.5 | | | 0.1 | 1.7 |
| NC RF acceleration | 104 | 0.1 | 4.1 | | | | 4.2 |
| SC RF linac | 140 | 0.1 | 3.4 | | | | 3.5 |
| SC RF RLAs | 10400 | 9.1 | 19.5 | | | | 28.6 |
| SC RF RCSs | 12566 | 11.3 | 11.8 | | | | 23.1 |
| Collider ring | 2600 | 2.3 | | 3.0 | 10 | | 15.3 |
| Totals | 26777 | 24.6 | 52.5 | 18.0 | 21.7 | 8.8 | 146 |

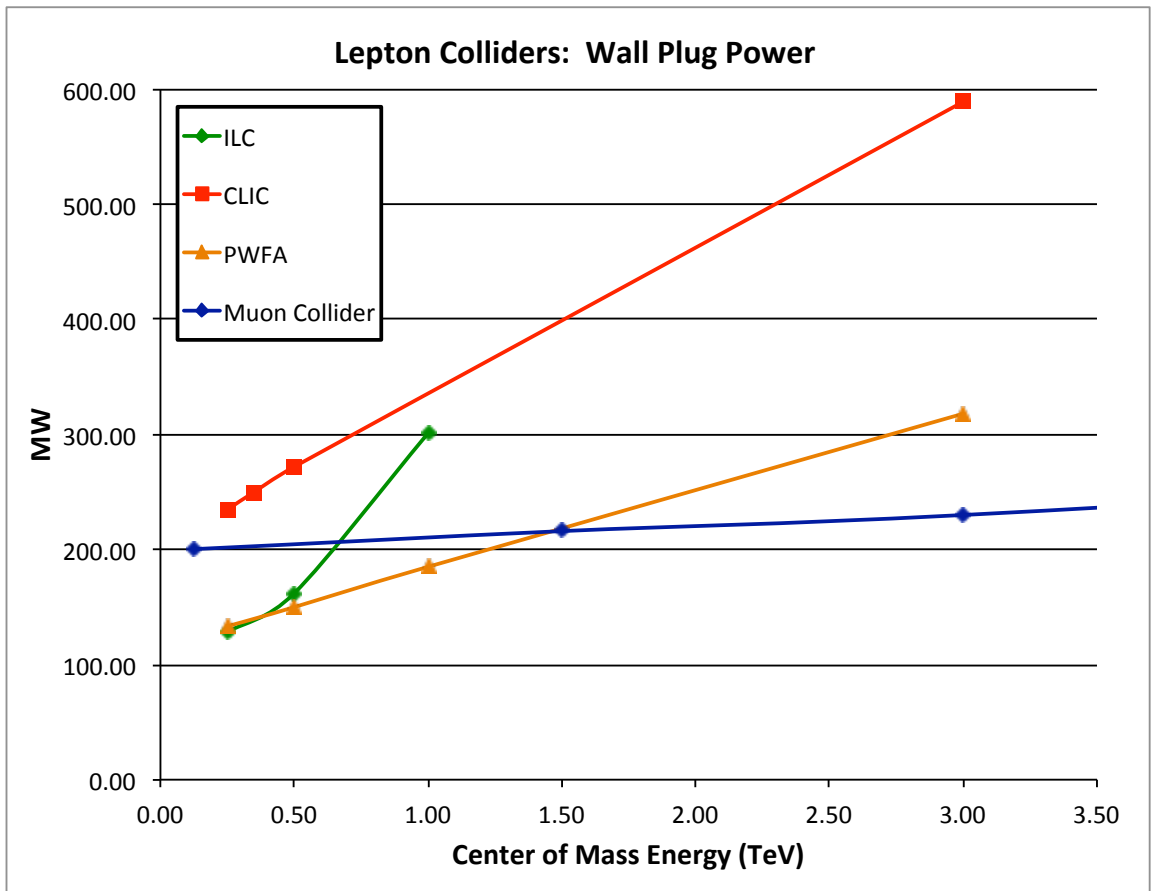

**Figure A-3**:  Wall-plug power evolution with colliding-beam energy

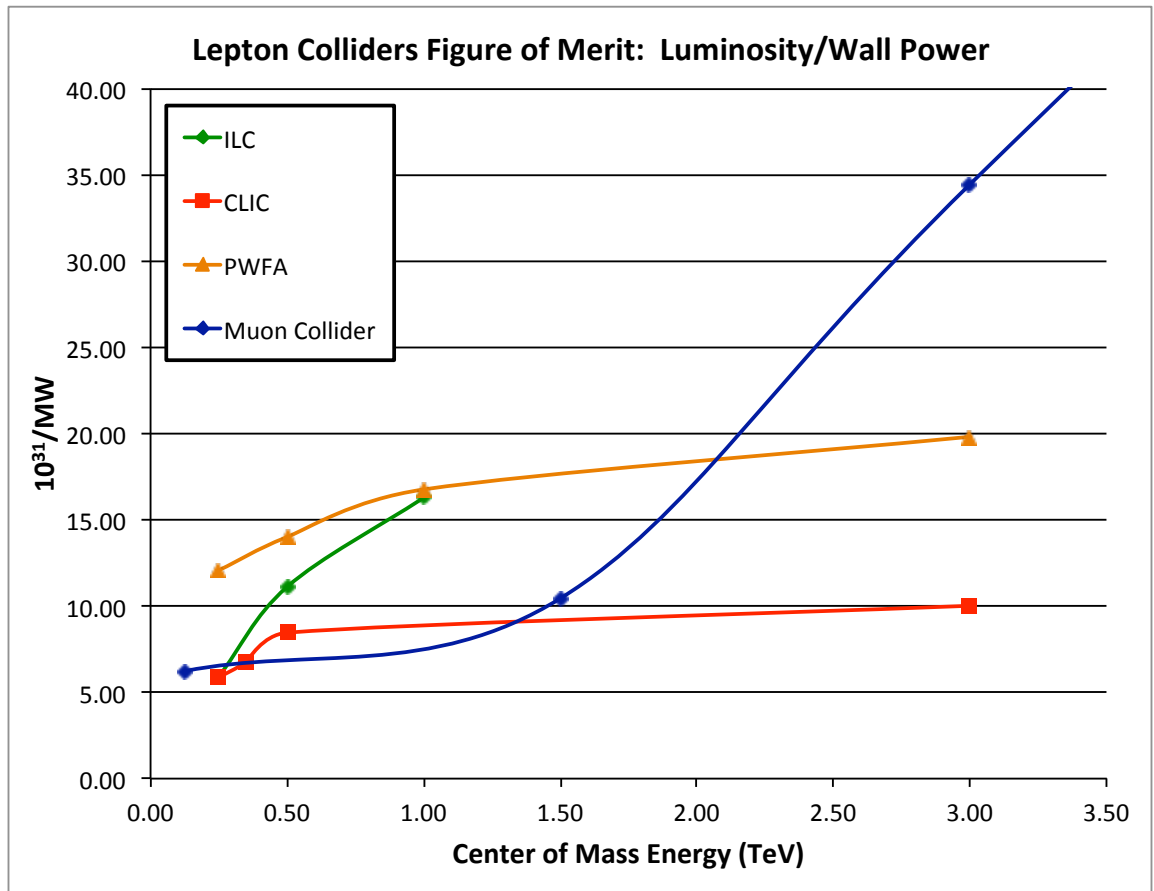

**Figure A-4:** Figure of merit: peak luminosity (within 1% colliding energy) normalized to wall-plug power

## A-5: R&D and Footprint of the Facility

*What are the critical technical challenges for muon colliders as a function of collider energy? What R&D must be done to address them and what are key demonstrations and milestones? What is the timescale and what are the pacing factors? What infrastructure is required and what is the characteristic footprint?*

R&D for the Muon Collider generally falls into two categories:

- Novel technologies unique to Muon Colliders;
- Conventional technologies where the operating parameters exceed the present state of the art.

The critical challenges include:

- A high-power proton linac and target station (up to 4 MW) although full power capability is not required for initial Neutrino or Higgs Factory operation;
- A 15–20 T capture solenoid;
- RF accelerating gradient in low frequency (325–975 MHz) structures immersed in high magnetic field as required for the front end and ionization cooling sections;
- Ionization cooling by 6 orders of magnitude (2 in each transverse plane and 2 in longitudinal plane);

# U.S. Muon Accelerator Program

- Very high field (> 30 T) solenoids utilizing high temperature superconducting (HTS) coils (only required for the multi-TeV collider final cooling section);
- Recirculating linacs (RLA) and rapid-cycling synchrotron (RCS) or fixed-field alternating-gradient (FFAG) rings for fast beam acceleration;
- A collider ring design and machine–detector interface (MDI) including absorbers for the decay products of the muon beams;
- Detector operation in a unique background environment caused by the muon decays around the ring.

The feasibility of each of these issues is being addressed by the Muon Accelerator Program (MAP). The MAP Feasibility Assessment is expected to conclude by the end of this decade. It will include the demonstration of the principle of ionization cooling in the MICE experiment at RAL with Step IV (utilizing absorbers without re-acceleration) experimental results expected starting in 2015 and with Step VI (utilizing a full cooling cell including re-acceleration) results anticipated by the end of the decade. Following MICE, nuSTORM could provide reasonably intense muon beams ($10^{10}$/pulse) for initial testing of the 6D cooling systems required for collider applications in time for an informed decision on moving forward with a Muon Collider project by the middle of the next decade.

The RLA and FFAG concepts are subjects of specific test facilities, respectively JEMMRLA as proposed at JLAB and EMMA in operation at Daresbury Laboratory in the UK.

A staged approach for muon accelerator facilities at FNAL has been developed which can support physics output at each stage along with validation of the systems required for subsequent stages. Such an approach minimizes the need for dedicated, standalone test facilities. The staging, as presently envisioned, consists of:

- nuSTORM validating cooling at reasonable intensity while providing important neutrino physics measurements.
- A lower-luminosity Neutrino Factory, NuMAX, powered by a proton beam with a reasonable power of 1 MW at 3 GeV as provided by Project X Stage IIa and requiring no muon cooling. It would be used as a powerful long-baseline neutrino source and serve as an R&D platform to test and validate transverse cooling (4D) at full muon bunch intensity ($10^{12}$/bunch) as required by a full-luminosity Neutrino Factory. In addition, it would validate the injector complex at the 1 MW level as well as the corresponding target, front end and 5 GeV RLA. Such a facility could be started around 2020.
- A high-luminosity Neutrino Factory (NuMAX+) upgraded from NuMAX by increasing the proton driver to the full power of 3 MW at 3 GeV provided by Project X Stage IIb, and ultimately to the nominal power of 4 MW at 8 GeV provided by Project X Stage IV with the corresponding target station upgrade. Performance would benefit from the 4D cooling validated by R&D at NuMAX. This facility does not require any longitudinal cooling but would be used as a muon source and an R&D platform to test and validate transverse and longitudinal cooling (6D) to full specification and nominal muon bunch intensity ($10^{12}$/bunch) as required by Muon Colliders.
- A low-energy Muon Collider (Higgs Factory) using the proton driver, target, front end and first stage acceleration of the Neutrino Factory as well as the 6D cooling previously developed but without final cooling. It would be upgraded in energy by further RLA or RCS. It could be used as a muon source and an R&D platform to test and validate the final cooling to small transverse emittances at nominal muon bunch intensity ($10^{12}$/bunch) as required by a high-energy Muon Collider. Such a facility could be launched by 2025.

- A high-energy Muon Collider in the TeV or multi-TeV energy range using all technologies tested and validated during the previous phases upgraded in energy by installing more RLA and/or RCS and a corresponding collider ring.

A complex integrating all above facilities in a staged approach would integrate well with Project X on the FNAL site as shown in Figure 2 of the main body.

### A-6: Cost Drivers

*What are the anticipated cost drivers in the research program? What are the major cost drivers for the collider facility?*

The principal cost driver of the R&D program is the demonstration of ionization cooling which is the most novel capability required for a muon collider. A dedicated test facility, the Muon Ionization Cooling Experiment (MICE), is being built and commissioned at Rutherford Appleton Laboratory (RAL) by an international collaboration.

Major cost drivers for the muon collider facility will include the proton driver and high-power target systems, the cooling system and the acceleration systems. The collider ring with its small circumference at low energy (i.e., a Higgs Factory) is not expected to be a cost driver, but its cost become more significant at high energy (TeV range). The cost of a muon collider is therefore expected to increase somewhat linearly with energy with a large offset due to the necessary systems to produce and cool the muon beam before acceleration, namely the proton driver, front end and cooling systems.

### A-7: Technology Applications

*Are there technology applications beyond energy frontier science that motivate development?*

Muon Collider and Neutrino Factory technology development is highly synergistic. In particular, there are overlapping needs for muon production, capture, cooling and acceleration. Joint development in these areas can enable a physics program spanning both the Energy and Intensity Frontiers. Furthermore, cooled muon beams also have potential applications in low- and medium-energy muon physics, as well as for homeland security and medicine.

Finally, the final cooling channel for a multi-TeV Muon Collider requires development of high temperature superconducting (HTS) solenoid magnets, and various systems in the collider facility would benefit from the improved performance that would be enabled by the development of very high field HTS accelerator magnets with good field quality. The development of this new class of conductors and magnets has potentially major impacts for applications outside high energy physics, in particular applications in science and the energy industry.

# U.S. Muon Accelerator Program

## Appendix B:  Projected Collider Parameters for Options Not Presently Included in the MAP Baseline Configurations

### B-1: Machine Parameters at the Top Quark Production Threshold

As part of the Snowmass process, the question was posed what parameters would be achievable at "intermediate" energies with a Muon Collider, i.e., those energies between the Higgs resonance and the TeV-scale, such as the top threshold.  This question was specifically aimed at understanding what potential trade-offs exist between energy resolution and luminosity as well as the overall performance that could be expected.  Because of the ability with the Muon collider to trade off transverse versus longitudinal beam emittance by picking different stopping points in the ionization cooling channel, significant optimization of these parameters is possible.  Table B-1 shows two sets of operating parameters at the top threshold.  The first column describes a configuration optimized to provide a center-of-mass energy resolution of 25 MeV with a luminosity of $7 \times 10^{32}\,\mathrm{cm^{-2}\,s^{-1}}$, while the second column describes a configuration optimized for luminosity at this energy.  In this case the center-of-mass energy resolution obtained is 250 MeV with a luminosity of $6 \times 10^{33}\,\mathrm{cm^{-2}\,s^{-1}}$.

**Table B-1:**  Two parameter sets for a Muon Collider at the top production threshold, one optimized for energy resolution and the other optimized for luminosity performance

| Muon Collider Parameters | | | |
|---|---|---|---|
| | | **Top Threshold Options** | |
| *Parameter* | *Units* | *High Resolution* | *High Luminosity* |
| CoM Energy | TeV | 0.35 | 0.35 |
| Avg. Luminosity | $10^{34}\mathrm{cm^{-2}s^{-1}}$ | 0.07 | 0.6 |
| Beam Energy Spread | % | 0.01 | 0.1 |
| Top Production/$10^7$sec | | 7,000[+] | 60,000[+] |
| Circumference | km | 0.7 | 0.7 |
| No. of IPs | | 1 | 1 |
| Repetition Rate | Hz | 15 | 15 |
| $\beta^*$ | cm | 1.5 | 0.5 |
| No. muons/bunch | $10^{12}$ | 4 | 3 |
| No. bunches/beam | | 1 | 1 |
| Norm. Trans. Emittance, $\varepsilon_{TN}$ | $\pi$ mm-rad | 0.2 | 0.05 |
| Norm. Long. Emittance, $\varepsilon_{LN}$ | $\pi$ mm-rad | 1.5 | 10 |
| Bunch Length, $\sigma_s$ | cm | 0.9 | 0.5 |
| Beam Size @ IP | $\mu$m | 43 | 13 |
| Beam-beam Parameter / IP | | 0.02 | 0.06 |
| Proton Driver Power | MW | 4 | 4 |

## B-2: An Overview of Potential Operating Energies for a Muon Collider

The range of center-of-mass energies that can be explored with a Muon Collider is of particular interest. The technology becomes interesting around the ~100 GeV energy scale where $e^+e^-$ circular colliders become challenging due to the large amount of synchrotron radiation emitted by the beam. Muon beams face a different radiation issue that potentially sets a maximum reasonable energy for a collider. As the muons decay in a collider ring, the resulting neutrinos are produced in a fan that will eventually break the earth's surface[42]. In order to keep the neutrino radiation within acceptable limits, present designs will likely be restricted to ≤ 10 TeV maximum center-of-mass energy. Table B-2 summarizes a range of machine parameters from the Higgs resonance up to 6 TeV.

**Table B-2:** A summary of potential parameters for muon colliders with center of mass energies ranging from the Higgs resonance up to 6 TeV.

| Parameter | Units | Higgs Factory | | Top Threshold Options | | Multi-TeV Baselines | | Accounts for Site Radiation Mitigation |
|---|---|---|---|---|---|---|---|---|
| | | Startup Operation | Production Operation | High Resolution | High Luminosity | | | |
| CoM Energy | TeV | 0.126 | 0.126 | 0.35 | 0.35 | 1.5 | 3.0 | 6.0 |
| Avg. Luminosity | $10^{34}$cm$^{-2}$s$^{-1}$ | 0.0017 | 0.008 | 0.07 | 0.6 | 1.25 | 4.4 | 12 |
| Beam Energy Spread | % | 0.003 | 0.004 | 0.01 | 0.1 | 0.1 | 0.1 | 0.1 |
| Higgs* or Top* Production/$10^7$sec | | 3,500* | 13,500* | 7,000* | 60,000* | 37,500* | 200,000* | 820,000* |
| Circumference | km | 0.3 | 0.3 | 0.7 | 0.7 | 2.5 | 4.5 | 6 |
| No. of IPs | | 1 | 1 | 1 | 1 | 2 | 2 | 2 |
| Repetition Rate | Hz | 30 | 15 | 15 | 15 | 15 | 12 | 6 |
| β* | cm | 3.3 | 1.7 | 1.5 | 0.5 | 1 (0.5-2) | 0.5 (0.3-3) | 0.25 |
| No. muons/bunch | $10^{12}$ | 2 | 4 | 4 | 3 | 2 | 2 | 2 |
| No. bunches/beam | | 1 | 1 | 1 | 1 | 1 | 1 | 1 |
| Norm. Trans. Emittance, $\varepsilon_{TN}$ | π mm-rad | 0.4 | 0.2 | 0.2 | 0.05 | 0.025 | 0.025 | 0.025 |
| Norm. Long. Emittance, $\varepsilon_{LN}$ | π mm-rad | 1 | 1.5 | 1.5 | 10 | 70 | 70 | 70 |
| Bunch Length, $\sigma_s$ | cm | 5.6 | 6.3 | 0.9 | 0.5 | 1 | 0.5 | 0.2 |
| Proton Driver Power | MW | 4[#] | 4 | 4 | 4 | 4 | 4 | 1.6 |

[#] Could begin operation with Project X Stage II beam

---

[42] R. Palmer, "An Overview of Muon Colliders," *ICFA Beam Dynamics Newsletter* **55** (2011), pp. 22–53.